\documentclass[aps,prb,twocolumn]{revtex4-2}
\makeatletter\if@twocolumn\PassOptionsToPackage{switch}{lineno}\else\fi\makeatother

\usepackage{tabulary,graphicx,amsfonts,amsmath,amssymb,amsbsy,dcolumn,bm}
\usepackage[utf8]{inputenc}
\usepackage[T1]{fontenc}

\usepackage{url,multirow,morefloats,floatflt,cancel,tfrupee}
\makeatletter

\AtBeginDocument{\@ifpackageloaded{textcomp}{}{\usepackage{textcomp}}}
\makeatother
\usepackage{colortbl}
\usepackage{xcolor}
\usepackage{pifont}
\usepackage[nointegrals]{wasysym}
\makeatletter

\def\mcWidth#1{\csname TY@F#1\endcsname+\tabcolsep}

\def\cAlignHack{\rightskip\@flushglue\leftskip\@flushglue\parindent\z@\parfillskip\z@skip}
\def\rAlignHack{\rightskip\z@skip\leftskip\@flushglue \parindent\z@\parfillskip\z@skip}

\@ifundefined{etal}{}{}

\usepackage{ifxetex}
\ifxetex\else\if@twocolumn\@ifpackageloaded{stfloats}{}{\usepackage{dblfloatfix}}\fi\fi

\AtBeginDocument{
\expandafter\ifx\csname eqalign\endcsname\relax
\def\eqalign#1{\null\vcenter{\def\\{\cr}\openup\jot\m@th
  \ialign{\strut$\displaystyle{##}$\hfil&$\displaystyle{{}##}$\hfil
      \crcr#1\crcr}}\,}
\fi
}

\AtBeginDocument{%
  \@ifpackageloaded{endfloat}%
   {\renewcommand\efloat@iwrite[1]{\immediate\expandafter\protected@write\csname efloat@post#1\endcsname{}}}{\newif\ifefloat@tables}%
}%

\def\BreakURLText#1{\@tfor\brk@tempa:=#1\do{\brk@tempa\hskip0pt}}
\let\lt=<
\let\gt=>
\def\processVert{\ifmmode|\else\textbar\fi}

\@ifundefined{subparagraph}{
\def\subparagraph{\@startsection{paragraph}{5}{2\parindent}{0ex plus 0.1ex minus 0.1ex}%
{0ex}{\normalfont\small\itshape}}%
}{}

\newcommand\role[1]{\unskip}
\newcommand\aucollab[1]{\unskip}
  
\@ifundefined{tsGraphicsScaleX}{\gdef\tsGraphicsScaleX{1}}{}
\@ifundefined{tsGraphicsScaleY}{\gdef\tsGraphicsScaleY{.9}}{}
\def\checkGraphicsWidth{\ifdim\Gin@nat@width>\linewidth
	\tsGraphicsScaleX\linewidth\else\Gin@nat@width\fi}

\def\checkGraphicsHeight{\ifdim\Gin@nat@height>.9\textheight
	\tsGraphicsScaleY\textheight\else\Gin@nat@height\fi}

\def\fixFloatSize#1{}
\let\ts@includegraphics\includegraphics

\def\inlinegraphic[#1]#2{{\edef\@tempa{#1}\edef\baseline@shift{\ifx\@tempa\@empty0\else#1\fi}\edef\tempZ{\the\numexpr(\numexpr(\baseline@shift*\f@size/100))}\protect\raisebox{\tempZ pt}{\ts@includegraphics{#2}}}}

\AtBeginDocument{\def\includegraphics{\@ifnextchar[{\ts@includegraphics}{\ts@includegraphics[width=\checkGraphicsWidth,height=\checkGraphicsHeight,keepaspectratio]}}}

\DeclareMathAlphabet{\mathpzc}{OT1}{pzc}{m}{it}

\def\URL#1#2{\@ifundefined{href}{#2}{\href{#1}{#2}}}

\def\UrlOrds{\do\*\do\-\do\~\do\'\do\"\do\-}%
\g@addto@macro{\UrlBreaks}{\UrlOrds}

\edef\fntEncoding{\f@encoding}

\makeatother

\newif\ifmultipleabstract\multipleabstractfalse%
\begin{document}

\title{The Continuum Model for Uniaxially Strained Bilayer Graphene Moir\'e Systems}
\author{Tong Liu$^{1, 3}$}
\author{X. R. Wang$^{2*}$}
\author{Jiansheng Wu$^{3*}$}
\affiliation{ $^{1}$Department of Physics\unskip, Hong Kong University of Science and Technology, Hong Kong }
\affiliation{ $^{2}$School of Science and Engineering, Chinese University of Hong Kong (Shenzhen), Shenzhen, 51817, China }
\affiliation{ $^{3}$Shenzhen Institute for Quantum Science and Engineering\unskip, Southern University of Science and Technology, 518055, Shenzhen, P. R. China}
\email[Corresponding author: ]{phxwan@cuhk.edu.cn (X.W.); jwu@sustech.edu.cn (J. W.)}



\begin{abstract}
We construct a continuum model for a one-dimensional moir\'e superlattice formed by stretching one layer of AB-stacked bilayer graphene along the $x$ direction by a factor $s$. Following the spirit of the Bistritzer--MacDonald model for twisted bilayer graphene, we treat the interlayer coupling as hopping between several Dirac points. At a critical stretch factor $s\approx1.018$ the two bands near the Fermi level touch, forming two degeneracy points along the $k_y$ direction. This gap closing is accompanied by a topological phase transition, in which the Chern number changes from $1$ to $-1$, and by a sign change of the Berry-curvature dipole, which we propose can be detected through the nonlinear Hall effect. We find that uniaxial strain modulates inter-Dirac-valley coupling, which drives band gap collapse and subsequent topological number inversion. This opens a route to engineer topological transport and quantum anomalous Hall effects via strain engineering of moir\'e heterostructures.
\end{abstract}\def\keywordstitle{Keywords}

\maketitle 
    
\section{Introduction}
\label{sec:intro}
Twisted bilayer graphene has become a groundbreaking platform for exploring strongly correlated electron physics and exotic topological quantum states~\cite{Serlin2020}, triggering widespread research enthusiasm in condensed matter physics. Its emergent flat moir\'e bands drastically suppress electron kinetic energy, enabling dominant electron-electron Coulomb interactions that give rise to intriguing phenomena such as correlated insulating states and unconventional superconductivity~\cite{Cao2018a,Cao2018b}, which not only deepen the fundamental understanding of correlated and topological quantum mechanisms but also offer promising prospects for next-generation low-power quantum electronic and optoelectronic devices. As a complementary and flexible alternative to twist engineering, strain-modulated bilayer graphene provides a distinctive strategy to construct moir\'e superlattices without relying on twist angles, enabling precise and continuous modulation of interlayer stacking configurations and low-energy electronic properties for advanced quantum device design. Existing theoretical investigations on biaxially strained bilayer graphene generally adopt the classic Bistritzer-MacDonald (BM) model, yet this approach suffers from an inherent limitation: it breaks the unidirectional lattice periodicity preserved in uniaxial strain systems, thus fails to capture the unique one-dimensional moir\'e characteristics and domain-level tunability~\cite{SanJose2013,Huang2018}. In contrast, uniaxial heterostrain can effectively generate robust one-dimensional moir\'e superlattices~\cite{Huder2018}, exhibiting unparalleled advantages in selective stacking configuration control and dimensional regulation of electronic states, which holds great potential for developing reconfigurable low-dimensional quantum devices and tunable topological systems. Nevertheless, current studies on uniaxially strained graphene moir\'e structures mainly depend on numerical simulations and first-principles calculations, while systematic analytical models are still absent. This shortage fails to explicitly reveal the quantitative relationships between structural strain parameters and electronic performances, hinders in-depth physical interpretation of strain-induced quantum modulation, and greatly restricts the rational design and practical application of corresponding graphene-based quantum devices. The purpose of this article is to construct an analytical model which can capture the main structure of energy bands of uniaxially strained graphene moir\'e structures, i.e. the one-Dimensional Moir\'e Superlattice.



As is well known, twisted bilayer graphene (TBG)  hosts a moir\'e superlattice, and its low-energy physics can be captured by the Bistritzer--MacDonald continuum model~\cite{LopesdosSantos2012}, which has been used to determine the magic angle~\cite{BistritzerMacDonald,Watson2023} and the Berry curvature~\cite{Liu2019,Zhang2019}. For the stretched graphene bilayer, the Dirac--Harper model provides another continuum description, which has been used to study its anomalous electrodynamics and quantum geometry~\cite{Timmel2020,Timmel2021}. In this paper we build a continuum model for the stretched bilayer graphene that is analogous to the Bistritzer--MacDonald model, and use it to describe phenomena that have been predicted. DFT calculations show that at a special stretch factor $s\approx1.018$ the two bands near the Fermi level close~\cite{Su2025}. Our model reproduces this special stretch factor and gives the positions of the degeneracy points. It also explains how the degeneracy forms and shows that the closing of the energy gap changes the topological number, which is proposed to be 
detected by the nonlinear Hall effect~\cite{Ma2019,Kang2019,Xu2018}. 


The remainder of this paper is structured as follows. Section \ref{sec:model} presents the theoretical model and basic formalism. Section \ref{sec:degeneracy} analyzes the characteristics of degeneracy points. The topological invariants are calculated and the experiment
to detect the topological phase transition is proposed in Section \ref{sec:topology}. Finally, Section \ref{sec:conclusion} summarizes the overall results and draws the conclusions.

\section{CONTINUUM MODEL FOR UNIAXIALLY STRETCHED BILAYER GRAPHENE}\label{sec:model}

\begin{figure}[htbp]
	\centering
	\includegraphics[width=\columnwidth]{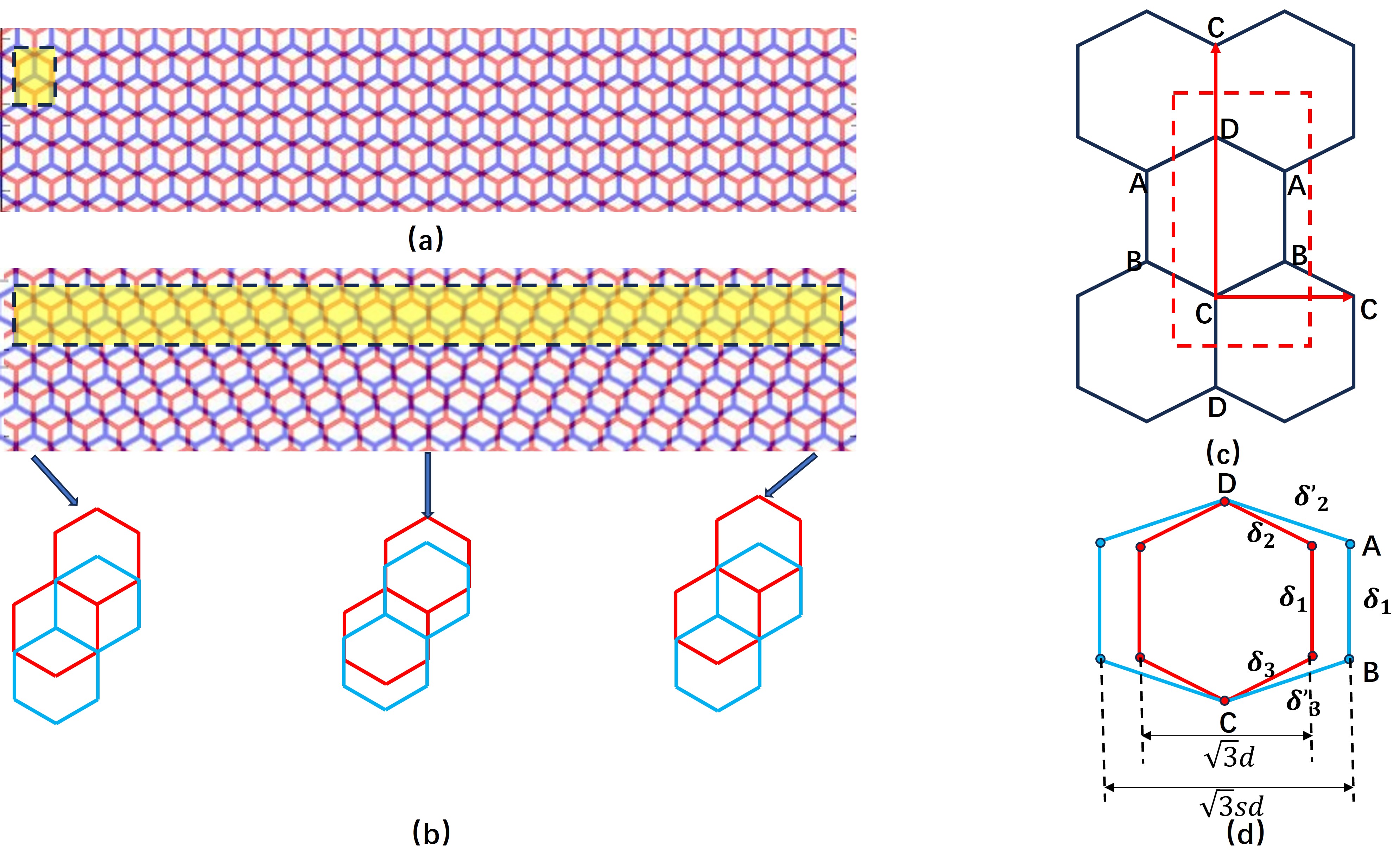}\\
	\caption{The structure in real space. (a) The AB-stacked bilayer graphene before stretching; the yellow region inside the black dashed line is the unit cell. (b) The stretched-bilayer structure studied in this paper, where the blue layer is the stretched one. The yellow region is the superlattice, which contains neither AA- nor BA-stacked regions. (c) The larger unit cell we choose in the single layer, containing four atoms $A$, $B$, $C$, and $D$; the red dashed line marks the unit cell. (d) How the stretching changes the unit cell, where $s$ is the stretch factor.}
	\label{Fig1}
\end{figure}

We consider  AB-stacked bilayer graphene and one layer is stretched.
If  a single graphene layer is stretched in one direction, both its geometry and its hopping amplitudes change. Here we stretch the graphene layer along the $x$ axis, as shown in Fig.~\ref{Fig1}, and define the stretch factor $s$ as the ratio of the basis-vector length after stretching to that before stretching. The hopping between two carbon atoms whose distance is changed becomes $t'=\Delta(s)\,t$, where $t=2.8$~eV is the unstrained nearest-neighbor hopping. The stretching also changes the reciprocal-space structure and the position of the $K$ valley.

For an AB-stacked bilayer graphene with one stretched layer, a new period forms along the stretching direction, whereas the periodicity perpendicular to it remains unchanged. This is a one-dimensional moir\'e superlattice. The period along the stretching direction is much larger than the original one when $s-1$ is small (typically $s-1$ is no larger than two percents), so the first Brillouin zone of the superlattice becomes greatly compressed along one reciprocal-space
direction. Taking the stretching direction as the $x$ axis, the Brillouin zone is reduced along $k_x$. In contrast to twisted bilayer graphene, the structure of this 1D moir\'e pattern depends on how the graphene layer is stretched; the structure studied here is shown in Fig.~\ref{Fig1}, and it contains neither AA- nor BA-stacked regions.

In this paper we refer to this 1D moir\'e superlattice structure as the stretched bilayer graphene, abbreviated as SBLG. The Hamiltonian of the SBLG system contains three parts:
\begin{equation}
H_{SBLG}=H_{\alpha}+H_{\beta}+H_{\alpha,\beta}
\label{eq:HSBLG}
\end{equation}
Here the first two parts are the single-layer Hamiltonians for the upper and lower layer respectively and the third part is the interlayer coupling term. We label the two layers as the $\alpha$ layer and the $\beta$ layer; the $\beta$ layer is the stretched along $x$ axis. Usually, for a single layer, we use the two-sublattice model, because the honeycomb unit cell contains two atoms. For the SBLG system, however, we need a larger unit cell. For the AB-stacked bilayer before stretching, the first Brillouin zone is a hexagon, while after stretching it becomes a rectangle. The two basis vectors of the superlattice in real space are $\mathbf{a}_{1}^{\mathrm{super}}=(0,3d)$ and $\mathbf{a}_{2}^{\mathrm{super}}=(\frac{s\sqrt{3}d}{s-1},0)$, where $d=1.42\times10^{-10}$~m is the nearest-neighbor distance in one layer. The corresponding reciprocal basis vectors are 
$\mathbf{b}_{1}^{\mathrm{super}}=(0,\frac{\sqrt{3}}{2}K)$ and $\mathbf{b}_{2}^{\mathrm{super}}=(\frac{3}{2}\frac{s-1}{s}K,0)$ with $K=\frac{4\sqrt{3}\pi}{9d}$, . Since the stretching occurs only in one direction, it does not affect the perpendicular direction, in both real and reciprocal space. We therefore use a larger unit cell containing four atoms, with real-space basis vectors $\mathbf{a}_{1}=(0,3d)$ and $\mathbf{a}_{2}=(\sqrt{3}d,0)$. The unit cell of the AB-stacked bilayer has the same size as that of the single layer; comparing it with the superlattice unit cell, we see that the stretching only changes the basis vector along the $x$ axis.

\begin{figure}[htbp]
	\centering
	\includegraphics[width=\columnwidth]{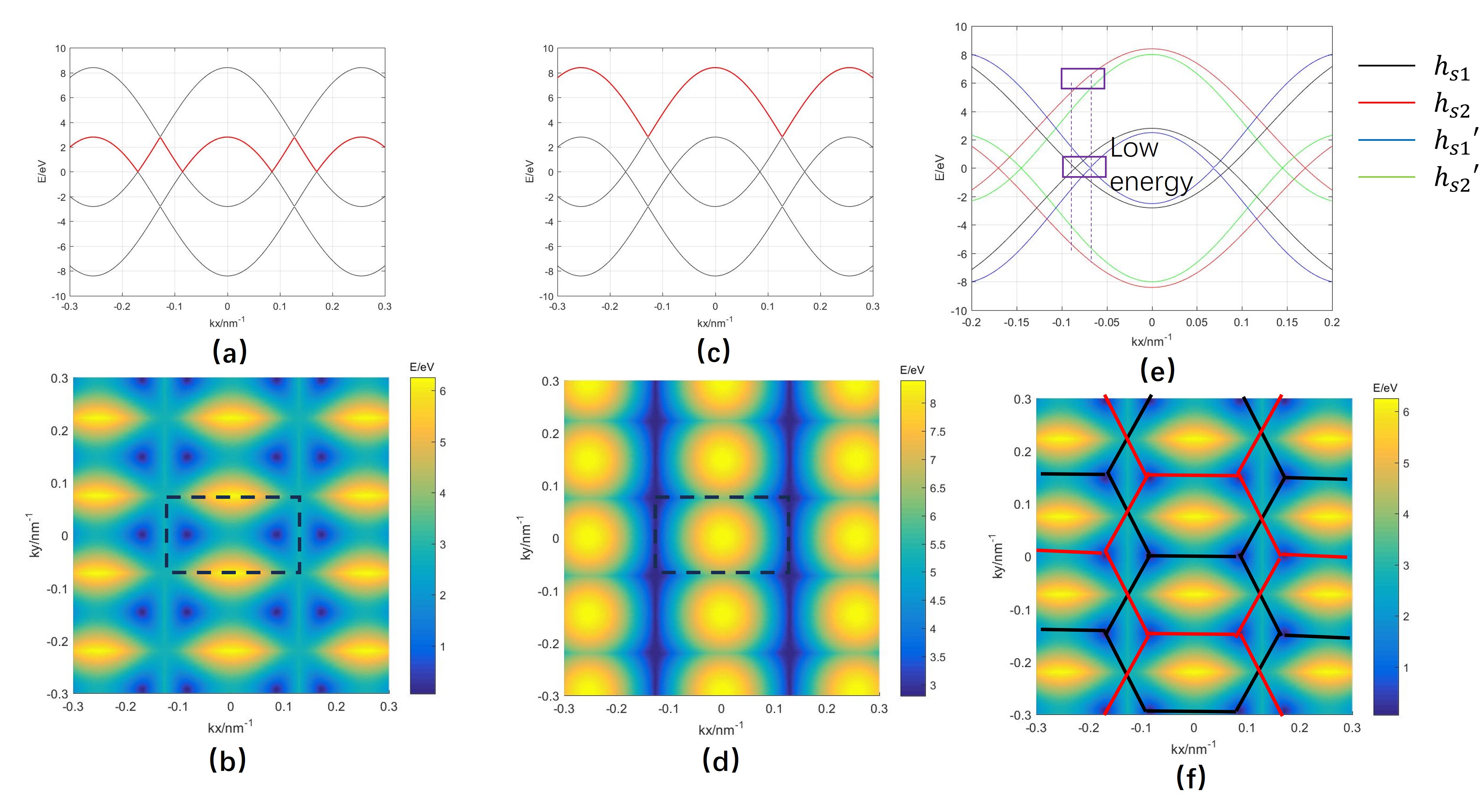}\\
	\caption{(a)--(d) Energy band structure given by Eq.~\eqref{eq:hsingle}; (a) and (c) are the band structure at $k_y=0$, while (b) and (d) show the red band in (a) and (c), respectively. The black dashed box marks the first Brillouin zone. (e) Band structure of the stretched bilayer without the interlayer interaction; the eight bands are given by the four Hamiltonians $h_{si}$ and $h_{si}'$ after the change of basis. In our study we only consider the low-energy part, so near the Dirac cone we need only keep the interaction between $h_{s1}$ and $h_{s1}'$, because the other two bands have higher energy. (f) Honeycomb structure of $h_{s1}$ (black) and $h_{s1}'$ (red).}
	\label{Fig2}
\end{figure}

The tight-binding Hamiltonian of unstrained single-layer graphene is $H=\sum_{\mathbf{R},\bm{\delta},i,j}c_{i}^{\dagger}(\mathbf{R})h_{\bm{\delta}}^{ij}c_{j}(\mathbf{R}+\bm{\delta})$. We consider only nearest-neighbor hopping, so in the unit cell the vector $\bm{\delta}$ can only be $\bm{\delta}_{1}$, $\bm{\delta}_{2}$, or $\bm{\delta}_{3}$, and $i,j$ can be $A,B,C,D$. After the Fourier transform, the Hamiltonian in matrix form is
\begin{equation}
h_{\mathrm{single}}=\begin{bmatrix}
 0 & \gamma_{1} & 0 & \gamma_{2}\\
 \gamma_{1}^{\dagger} & 0 &  \gamma_{2}^{\dagger}  & 0\\
 0 &  \gamma_{2} & 0 & \gamma_{1}\\
\gamma_{2}^{\dagger} & 0 &\gamma_{1}^{\dagger} & 0
\end{bmatrix},
\label{eq:hsingle}
\end{equation}
where $\gamma_{1}=te^{i\mathbf{k}\cdot\bm{\delta}_{1}}$ and $\gamma_{2}=te^{i\mathbf{k}\cdot\bm{\delta}_{2}}+te^{i\mathbf{k}\cdot\bm{\delta}_{3}}$. The basis of this Hamiltonian is $\Psi=[C_{A}(\mathbf{k}),C_{B}(\mathbf{k}),C_{C}(\mathbf{k}),C_{D}(\mathbf{k})]$. We find that the first Brillouin zone is a rectangle with reciprocal vectors $\mathbf{G}_{1}=(0,\frac{\sqrt{3}}{2}K)$ and $\mathbf{G}_{2}=(\frac{3}{2}K,0)$, and that there are four energy bands. The band structure is shown in Fig.~\ref{Fig2}. We now define a new basis $\Psi_{X}=[X_{1}(\mathbf{k}),X_{2}(\mathbf{k}),X_{3}(\mathbf{k}),X_{4}(\mathbf{k})]$ with
\begin{equation}
\begin{aligned}
&X_{1}=C_{A}(\mathbf{k})+C_{C}(\mathbf{k}),\\
&X_{2}=C_{B}(\mathbf{k})+C_{D}(\mathbf{k}),\\
&X_{3}=C_{A}(\mathbf{k})-C_{C}(\mathbf{k}),\\
&X_{4}=C_{B}(\mathbf{k})-C_{D}(\mathbf{k}).
\end{aligned}
\label{eq:newbasis}
\end{equation}
In this basis the Hamiltonian becomes
\begin{equation}
\begin{aligned}
h_{\mathrm{single}}&=\begin{bmatrix}
 0 & \gamma_{1}+ \gamma_{2} & 0 &0\\
 \gamma_{1}^{\dagger}+ \gamma_{2}^{\dagger} & 0 & 0  & 0\\
 0 &  0 & 0 & \gamma_{1}-\gamma_{2}\\
0 & 0 &\gamma_{1}^{\dagger}-\gamma_{2}^{\dagger} & 0
\end{bmatrix}\\
&=\begin{bmatrix}
{h_{s1}} & 0\\
0 & {h_{s2}}
\end{bmatrix}.
\end{aligned}
\label{eq:hsingleX}
\end{equation}
Clearly $h_{s1}$ has the same form as the two-sublattice Hamiltonian of the single layer. It can also be shown that $h_{s2}(\mathbf{k})=h_{s1}(\mathbf{k}+(\frac{3}{2}K,0))=h_{s1}(\mathbf{k}+(0,\frac{\sqrt{3}}{2}K))$. Thus the four bands split into two groups, each having the same structure as the two-sublattice band of the single layer, as confirmed by the band structure in Fig.~\ref{Fig2}.

After the graphene layer is stretched as in Fig.~\ref{Fig1}(d), both the geometry and the hopping amplitudes change, and the Hamiltonian becomes
\begin{equation}
\begin{aligned}
h_{\mathrm{single}}'&=\begin{bmatrix}
 0 & \gamma_{1}+ \gamma_{2}' & 0 &0\\
 \gamma_{1}^{\dagger}+ {\gamma_{2}'}^{\dagger} & 0 & 0  & 0\\
 0 &  0 & 0 & \gamma_{1}-\gamma_{2}'\\
0 & 0 &\gamma_{1}^{\dagger}-{\gamma_{2}'}^{\dagger} & 0
\end{bmatrix}\\
&=\begin{bmatrix}
{h_{s1}}' & 0\\
0 & {h_{s2}}'
\end{bmatrix},
\end{aligned}
\label{eq:hsinglep}
\end{equation}
where $\gamma_{2}'=t'e^{i\mathbf{k}\cdot\bm{\delta}_{2}'}+t'e^{i\mathbf{k}\cdot\bm{\delta}_{3}'}$. The value of $t'$ depends on the distance between the two atoms, so we focus on $\bm{\delta}_{2}'$ and $\bm{\delta}_{3}'$. In our study the stretch factor $s$ is small (less than $1.02$), so $|\bm{\delta}_{2}|=|\bm{\delta}_{3}|=\sqrt{\frac{3}{4}s^{2}+\frac{1}{4}}\,d\approx sd$ and $t'=te^{-2.218(|\bm{\delta}_{2}|/d-1)}=te^{-2.218(s-1)}$. The stretching in the $x$ direction in real space compresses the $k_x$ direction in reciprocal space, so the reciprocal vectors become $\mathbf{G}_{1}'=(0,\frac{\sqrt{3}}{2}K)$ and $\mathbf{G}_{2}'=(\frac{3}{2s}K,0)$. In addition, the change of $t'$ shifts the Dirac cone away from the $K$ and $K'$ points. The distance between the Dirac cone and the $K$ point is $\Delta D=\frac{2\sqrt{3}}{3ds}\bigl[\frac{2\pi}{3}-\arccos(-\frac{t'}{2t})\bigr]$ along the $k_x$ axis.

Now consider the interlayer coupling term $H_{\alpha,\beta}$ in Eq.~\eqref{eq:HSBLG}. It is given by 
\begin{equation}
H_{\alpha,\beta}=\sum_{\mathbf{R}_{\alpha,i},\mathbf{R}_{\beta,j}}C_{\alpha,i}(\mathbf{R}_{\alpha,i})^{\dagger}\,t(|\mathbf{R}_{\alpha,i}-\mathbf{R}_{\beta,j}|)\,C_{\beta,j}(\mathbf{R}_{\beta,j}).
\label{eq:H-interlayer}
\end{equation}
Here $i$ and $j$ can be $A,B,C,D$. The distance between two atoms in different layers is $\sqrt{l^{2}+{d_{\perp}}^{2}}$, where $d_{\perp}$ is the interlayer distance and $l$ is the in-plane separation. We can usually ignore the interaction between atoms with $l$ larger than $d$. Table~\ref{Tab1} lists the values of $l$ between the atoms:
\begin{table}[htbp]
\caption{Values of the in-plane separation $l$ (in units of $d$) between atoms in different layers.}
\label{Tab1}
\begin{ruledtabular}
\begin{tabular}{c|c|c|c|c}
 & $\mathbf{\alpha A}$ & $\mathbf{\alpha B}$ & $\mathbf{\alpha C}$ & $\mathbf{\alpha D}$ \\
\hline
$\mathbf{\beta A}$ & $[d,\frac{\sqrt{7}}{2}d]$ & $[0,\frac{\sqrt{3}}{2}d]$ & $[\frac{1}{2}d,d]$ & $[\frac{3}{2}d,\sqrt{3}d]$ \\
\hline
$\mathbf{\beta B}$ & $[d,\frac{\sqrt{7}}{2}d]$ & $[d,\frac{\sqrt{7}}{2}d]$ & $[\frac{1}{2}d,d]$ & $[\frac{1}{2}d,d]$ \\
\hline
$\mathbf{\beta C}$ & $[\frac{1}{2}d,d]$ & $[\frac{3}{2}d,\sqrt{3}d]$ & $[d,\frac{\sqrt{7}}{2}d]$ & $[0,\frac{\sqrt{3}}{2}d]$ \\
\hline
$\mathbf{\beta D}$ & $[\frac{1}{2}d,d]$ & $[\frac{1}{2}d,d]$ & $[d,\frac{\sqrt{7}}{2}d]$ & $[d,\frac{\sqrt{7}}{2}d]$ \\
\end{tabular}
\end{ruledtabular}
\end{table}

Ignoring the terms with distance larger than $d$, the interlayer term becomes
\begin{equation}
H_{\alpha,\beta}=\Psi_{\alpha,C}\begin{bmatrix}
 0 &w_{1}& w_{2}&0\\
 0 & 0 & w_{3}  & w_{4}\\
 w_{5} &  0 & 0 & w_{6}\\
w_{7} & w_{8} & 0 & 0
\end{bmatrix}\Psi_{\beta,C}^{\dagger},
\label{eq:Hinter}
\end{equation}
where $w_{i}$ are the interlayer interaction factors, and the two basis vectors are $\Psi_{\alpha,C}=[C_{\alpha,A}(\mathbf{k}),C_{\alpha,B}(\mathbf{k}),C_{\alpha,C}(\mathbf{k}),C_{\alpha,D}(\mathbf{k})]$ and $\Psi_{\beta,C}=[C_{\beta,A}(\mathbf{k}),C_{\beta,B}(\mathbf{k}),C_{\beta,C}(\mathbf{k}),C_{\beta,D}(\mathbf{k})]$. Rewriting the Hamiltonian in the new basis $\Psi_{\alpha,X}=[X_{\alpha,1}(\mathbf{k}),X_{\alpha,2}(\mathbf{k}),X_{\alpha,3}(\mathbf{k}),X_{\alpha,4}(\mathbf{k})]$ and $\Psi_{\beta,X}=[X_{\beta,1}(\mathbf{k}),X_{\beta,2}(\mathbf{k}),X_{\beta,3}(\mathbf{k}),X_{\beta,4}(\mathbf{k})]$ defined above, the matrix part becomes
\begin{equation}
h_{\alpha,\beta}=\begin{bmatrix}
 w_{2} & w_{1} & 0 & 0\\
 w_{3} & w_{4} & 0  & 0\\
 0 &  0 & -w_{5} & w_{6}\\
0 & 0 & -w_{7} & -w_{8}
\end{bmatrix}=\begin{bmatrix}
T_{\alpha,\beta} & 0\\
0 & {T_{\alpha,\beta}}'
\end{bmatrix}.
\label{eq:hab}
\end{equation}
Clearly the matrix part can be decomposed into two $2\times2$ blocks. Using the basis $\Psi=[\Psi_{\alpha,X},\Psi_{\beta,X}]$, the matrix part of Eq.~\eqref{eq:HSBLG} is
\begin{equation}
h_{SBLG}=\begin{bmatrix}
h_{s1} & 0 & T_{\alpha,\beta} & 0\\
0 & h_{s2} & 0 & {T_{\alpha,\beta}}'\\
{T_{\alpha,\beta}}^{\dagger} & 0 & h_{s1}' & 0\\
0 & {T_{\alpha,\beta}}^{'\dagger} & 0 & h_{s2}'
\end{bmatrix}.
\label{eq:hSBLG}
\end{equation}
We then define a new basis $\Psi'=[\Psi_{1,X},\Psi_{2,X}]$ with $\Psi_{1,X}=[X_{\alpha,1}(\mathbf{k}),X_{\alpha,2}(\mathbf{k}),X_{\beta,1}(\mathbf{k}),X_{\beta,2}(\mathbf{k})]$ and $\Psi_{2,X}=[X_{\alpha,3}(\mathbf{k}),X_{\alpha,4}(\mathbf{k}),X_{\beta,3}(\mathbf{k}),X_{\beta,4}(\mathbf{k})]$, and rewrite Eq.~\eqref{eq:hSBLG} as
\begin{equation}
h_{SBLG}=\begin{bmatrix}
h_{s1} & T_{\alpha,\beta} & 0 & 0\\
{T_{\alpha,\beta}}^{\dagger} & h_{s1}' & 0 & 0\\
0 & 0 & h_{s2} & {T_{\alpha,\beta}}'\\
0 & 0 & {T_{\alpha,\beta}}^{'\dagger} & h_{s2}'
\end{bmatrix}=\begin{bmatrix}
h_{sg} & 0\\
0 & h_{sg}'
\end{bmatrix}.
\label{eq:hSBLGblock}
\end{equation}
Equation~\eqref{eq:hSBLGblock} shows that the total structure can be divided into two independent parts, each describing the interaction between two honeycomb structures, one of which has been stretched, as shown in Fig.~\ref{Fig2}. In our study we focus on the low-energy part. We note, however, that this model also contains part of the high-energy band structure, and that the corresponding interlayer interaction factor has the same order of magnitude as the low-energy one, differing only by a phase. Whether the conclusions of the Bistritzer--MacDonald continuum model can be extended to the high-energy region will be discussed in a future work. Ignoring the interlayer interaction gives the band structure in Fig.~\ref{Fig2}(e): near the Dirac cone there are only two low-energy bands, given by $h_{s1}$ and $h_{s1}'$. Equation~\eqref{eq:hSBLGblock} shows that the interaction between $h_{s1}$ and $h_{s1}'$ is independent of that between $h_{s2}$ and $h_{s2}'$, so we need only keep the $h_{sg}$ term in our model.

Near the Dirac cone we can rewrite the Hamiltonians $h_{s1}$ and $h_{s1}'$ as
\begin{equation}
\begin{aligned}
&h_{s1}(\mathbf{k})=v\begin{bmatrix}
0 & k_{x}+ik_{y}\\
k_{x}-ik_{y} & 0
\end{bmatrix},\\
&h_{s1}'(\mathbf{k})=\begin{bmatrix}
0 & v_{x}(k_{x}+\Delta D)+ivk_{y}\\
v_{x}(k_{x}+\Delta D)-ivk_{y} & 0
\end{bmatrix}.
\end{aligned}
\label{eq:hs1near}
\end{equation}
Here $v=\frac{3dt}{2}$ and $v_{x}=-\sqrt{3}dst'\sqrt{1-\frac{t^{2}}{4t'^{2}}}$; when the stretch factor $s$ is less than $1.02$, which is the range we mainly focus on, we can set $v_{Fx}=v_{F}$. Now we need to determine the interlayer hopping factors $w_{i}$ in Eq.~\eqref{eq:hab}. Since the model only contains $T_{\alpha,\beta}$, we only need to focus on $i=1,2,3,4$. The interlayer interaction in real space is $H_{\alpha,\beta}=\sum_{i=1,2;\,j=1,2;\,\mathbf{r}_{X,\alpha,i};\,\mathbf{r}_{X,\beta,j}}X_{\alpha,i}(\mathbf{r}_{X,\alpha,i})\,t(|\mathbf{r}_{X,\alpha,i}-\mathbf{r}_{X,\beta,j}|)\,X_{\beta,j}(\mathbf{r}_{X,\beta,j})^{\dagger}$. For the basis $\Psi_{X}$ we cannot obtain the position vectors $\mathbf{r}_{X}$ directly, but Eq.~\eqref{eq:Hinter} shows how to calculate the factors in the $\Psi_{C}$ basis. Following the same method used in analyzing the twisted-bilayer graphene system~\cite{Catarina2019}, we obtain the interlayer hopping term as
\begin{equation}
T_{1}=w_{0}\begin{bmatrix}
1  &1 \\
1  &1
\end{bmatrix}, \qquad
T_{2}=w_{0}\begin{bmatrix}
-1  & 2 \\
-1  &-1
\end{bmatrix}.
\label{eq:T1T2}
\end{equation}
The details of this calculation are given in the Supplemental Material. Together with the single-layer part, the total Hamiltonian of our model is
\begin{equation}
h_{SBLG}(\mathbf{k})=\begin{bmatrix}
 h_{s1}(\mathbf{k}) & T_{1} & T_{2}\\
 T^{\dagger}_{1} & h_{s1}'(\mathbf{k}-\mathbf{q}_{1}) & 0\\
 T^{\dagger}_{2} & 0 & h_{s1}'(\mathbf{k}-\mathbf{q}_{2})
\end{bmatrix},
\label{eq:hSBLGk}
\end{equation}
with $\mathbf{q}_{1}=-(\frac{s-1}{s}K+\Delta D,0)$ and $\mathbf{q}_{2}=(\frac{s-1}{2s}K-\Delta D,0)$.

This Hamiltonian shows that the states with momentum $\mathbf{k}$ near the $K$ valley of the $\alpha$ layer are directly coupled to two types of states with momenta $\mathbf{k}-\mathbf{q}_{1}$ and $\mathbf{k}-\mathbf{q}_{2}$ near the $K$ valley of the $\beta$ layer. In turn, these two types of states in the $\beta$ layer are directly coupled to three types of states with momenta $\mathbf{k}-\mathbf{q}_{1}+\mathbf{q}_{2}$, $\mathbf{k}$, and $\mathbf{k}+\mathbf{q}_{1}-\mathbf{q}_{2}$. To describe the full coupling between the two layers, we have to include more states from both layers, so the Hamiltonian grows larger and larger. The momenta of the states in the same layer follow the periodicity of $\mathbf{q}_{1}-\mathbf{q}_{2}$, so a large Hamiltonian is periodic in $\mathbf{q}_{1}-\mathbf{q}_{2}$ along the $k_x$ axis (note that $\mathbf{q}_{1}-\mathbf{q}_{2}$ is along the $k_x$ axis). This gives the period of the moir\'e pattern along the $k_x$ axis. For a more accurate numerical result, we can enlarge the Hamiltonian. We studied an $8\times8$ Hamiltonian expanded near $K_{\alpha,1}$ and $K_{\beta,1}$:
\begin{equation}
H_{8\times8}=\resizebox{0.78\columnwidth}{!}{$
\begin{bmatrix}
h_{s1}'(\mathbf{k}-\mathbf{q}_{1})  & T_{1}^{\dagger} & 0 & 0\\
T_{1}  & h_{s1}(\mathbf{k}) & T_{2} & 0\\
 0 & T_{2}^{\dagger} & h_{s1}'(\mathbf{k}-\mathbf{q}_{2}) & T_{1}^{\dagger}\\
 0 & 0 & T_{1}& h_{s1}(\mathbf{k}-\mathbf{q}_{2}+\mathbf{q}_{1})
\end{bmatrix}
$}.
\label{eq:H88}
\end{equation}
We find that at a special stretch factor $s$, two degeneracy points form at the Fermi level along the $k_y$ direction, consistent with the DFT result~\cite{Su2025}.
\begin{figure}[htbp]
	\centering
	\includegraphics[width=\columnwidth]{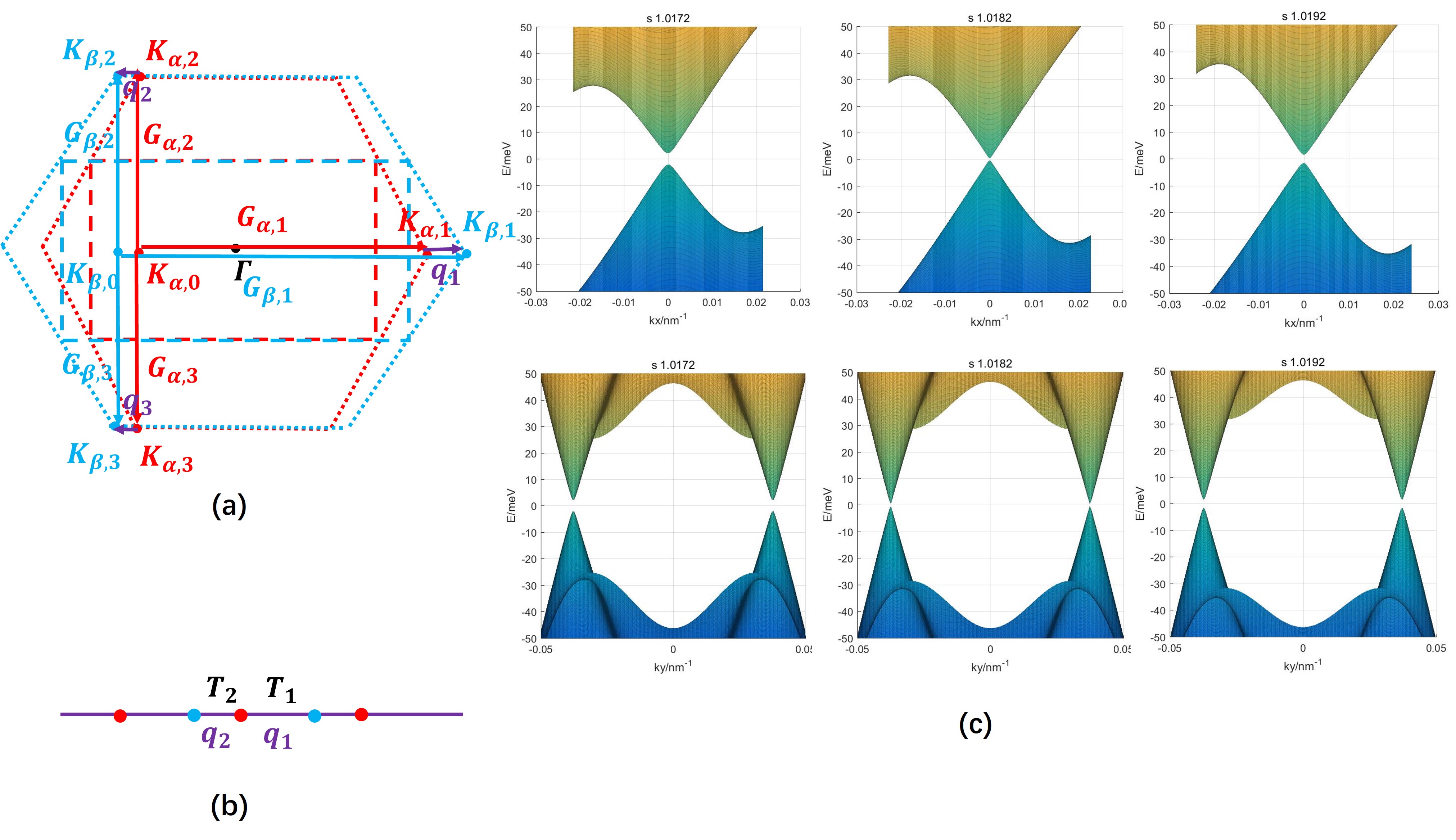}\\
	\caption{(a) The red dashed rectangle is the first Brillouin zone of the $\alpha$ layer and the blue dashed rectangle is that of the $\beta$ layer. By applying the three groups of reciprocal lattice vectors $\mathbf{G}_{l,i}$, we find that the three types of interaction can be regarded as hopping between three groups of $K_{l,i}$. For the three $K$ valleys from the same layer, we find that they belong to the same honeycomb structure $h_{s1}$ or $h_{s1}'$, shown by the dotted lines. This also proves the conclusion that the hopping between $h_{s1}$ and $h_{s1}'$ is independent. (b) There are two types of hopping from the $\alpha$ layer to the $\beta$ layer, and likewise from the $\beta$ layer to the $\alpha$ layer. Thus we can expand as in the TBLG system and obtain a 1D chain, i.e., the 1D moir\'e pattern along the $k_x$ axis. (c) Low-energy bands given by our model: at the stretch factor $s=1.0182$ the bands close at $E=0$~meV, and there are two degeneracy points along the $k_y$ direction.}
	\label{Fig3}
\end{figure}

It is important to emphasize that, although our previous calculations retained only the low-energy components, the first Brillouin zone of the two-dimensional rectangular superlattice should be rectangular geometry discribed by $\mathbf{b_{1}^{super}}$ and $\mathbf{b_{2}^{super}}$.However, the result of $h_{SBLG}$ given by $h_{sg}$ is periodic only along $\mathbf{b_{2}^{super}}$. The monolayer terms $h_{s1}$ and $h_{s1}'$ shift out of the low-energy window upon a translation of $\mathbf{b_{1}^{super}}$ from the Dirac point. In contrast, the terms $h_{s2}$ and $h_{s2}'$ come to reside in the vicinity of the Dirac cone. Our calculation in the supplemental material prove that the low-energy dispersion of $h_{sg}'$ coincides identically with that of $h_{sg}$. This result substantiates that the low-energy bands near the Fermi surface within the first Brillouin zone are jointly governed by both $h_{sg}$ and $h_{sg}'$.

 
\section{MECHANISM OF STRAIN-DRIVEN FERMI GAP CLOSURE AND ANALYTICAL SOLUTIONS OF DEGENERACY POINTS}\label{sec:degeneracy}

In this section we discuss how the degeneracy points form and at which stretch factor $s$ they appear.

Our model can be regarded as a coupling between Dirac points and is always captured by a simple $4\times4$ matrix:
\begin{equation}
H=\begin{bmatrix}
h(\mathbf{k}) & T \\
T^{\dagger}&h(\mathbf{k}+\mathbf{\Delta})
\end{bmatrix}=\begin{bmatrix}
h_{\alpha} & T \\
T^{\dagger}&h_{\beta}
\end{bmatrix},
\label{eq:Hdeg}
\end{equation}
where the $h_{l}$ are Dirac-cone Hamiltonians with off-diagonal terms $h_{l}^{12}=X_{l}-iY_{l}=h_{l}^{21\dagger}$. The relation $X_{\beta}=X_{\alpha}-v\Delta$ shows that the two Dirac points are not at the same location but are separated along the $k_x$ axis. The coupling matrix $T$ is given by
\begin{equation}
T=\begin{bmatrix}
a & b+c \\
b-c & a
\end{bmatrix},
\label{eq:T}
\end{equation}
where $a$, $b$, and $c$ are real numbers, because in $H_{SBLG}^{(1)}$ all the entries of the coupling matrix are real. By varying the four parameters $a$, $b$, $c$, and $\Delta$, we can determine in which situations degeneracy points appear and their locations. The influence of the parameters on the degeneracy points has also been discussed in the Dirac--Harper model~\cite{Timmel2021}; here, however, we adopt a different approach, converting the $4\times4$ matrix into a $2\times2$ one (the $M$-matrix method), in order to make the calculation more precise. Since it is easy to check whether a $2\times2$ Hamiltonian has degeneracy points, this conversion is particularly convenient.

The eigenvector of $H$ can be written as $\varphi=(\varphi_{1},\varphi_{2})$, where each $\varphi_{i}$ has two components. This gives two equations:
\begin{equation}
h_{\alpha}\varphi_{1}+T\varphi_{2}=E\varphi_{1},\qquad
T^{\dagger}\varphi_{1}+h_{\beta}\varphi_{2}=E\varphi_{2},
\label{eq:eig}
\end{equation}
which lead to
\begin{equation}
\Bigl[h_{\alpha}+\frac{1}{E^{2}-\gamma^{\dagger}\gamma}\bigl(ETT^{\dagger}+Th_{\beta}T^{\dagger}\bigr)\Bigr]\varphi_{1}=E\varphi_{1}.
\label{eq:Meff}
\end{equation}
The terms in square brackets can be regarded as a $2\times2$ matrix $M$. This equation then means that the $2\times2$ matrix $M$ has eigenvalue $E$ and eigenvector $\varphi_{1}$. The matrix elements of $M$ are
\begin{equation}
\begin{aligned}
&M_{11}=L_{\beta}\bigl[(a^{2}+(b+c)^{2})E+2a(b+c)X_{\beta}\bigr],\\
&M_{12}=L_{\beta}\bigl[2abE+a^{2}\gamma_{\beta}+(b^{2}-c^{2})\gamma_{\beta}^{\dagger}\bigr]+\gamma_{\alpha},\\
&M_{21}=L_{\beta}\bigl[2abE+a^{2}\gamma_{\beta}^{\dagger}+(b^{2}-c^{2})\gamma_{\beta}\bigr]+\gamma_{\alpha}^{\dagger},\\
&M_{22}=L_{\beta}\bigl[(a^{2}+(b-c)^{2})E+2a(b-c)X_{\beta}\bigr],
\end{aligned}
\label{eq:Melem}
\end{equation}
where $\gamma_{l}=X_{l}-iY_{l}$ and $L_{\beta}=\frac{1}{E^{2}-\gamma_{\beta}^{\dagger}\gamma_{\beta}}$. We take the position of the $\alpha$ Dirac point as the origin, so $X_{\alpha}=-vk_{x}$ and $Y_{\alpha}=-vk_{y}$. The matrix $M$ has a degeneracy point if and only if we can find a pair $(k_{x},k_{y})$ such that $M_{11}=M_{22}=E$ and $M_{12}=M_{21}=0$; this pair gives the location of the degeneracy point. The details of the numerical calculation are given in the Supplemental Material. For our stretched bilayer graphene model, the Dirac points lie along the $k_x$ axis and the two degeneracy points lie along the $k_y$ axis, which corresponds to the situation $\Delta=0$ with $a$, $b$, $c$ nonzero. Focusing on the coupling terms of our model, we find that for $T_{1}$: $a=w_{0}$, $b=w_{0}$, $c=0$; and for $T_{2}$: $a=w_{0}$, $b=0.5w_{0}$, $c=1.5w_{0}$. According to the analysis, $T_{1}$ moves the Dirac points along the $k_x$ axis by the distance
\begin{equation}
\left|\Delta_{x}\right|=\frac{\sqrt{v^{2}\Delta^{2}+16w_{0}^{2}}+v\Delta}{4v}-\frac{\Delta}{2v}.
\label{eq:dx}
\end{equation}
Considering the four points from left to right, labeled $\beta_{1}$, $\alpha_{1}$, $\beta_{2}$, and $\alpha_{2}$, the couplings between $\beta_{1},\alpha_{1}$ and between $\beta_{2},\alpha_{2}$ are both $T_{1}$, which moves $\alpha_{1}$ to the right and $\beta_{2}$ to the left. At a special factor $s$ (which affects the value of $\Delta$), $\alpha_{1}$ lies directly above $\beta_{2}$, and the band structure becomes similar to the case with $a\neq0$ and $b,c,\Delta=0$. On the other hand, the coupling $T_{2}$ between $\alpha_{1}$ and $\beta_{2}$ produces two degeneracy points along the $k_y$ axis. We therefore conclude that, to observe this phenomenon, the Hamiltonian should contain at least one structure with two $T_{1}$ couplings and one $T_{2}$ coupling, i.e., at least four points; a six-band model can produce this phenomenon. The special stretch factor is obtained by solving
\begin{equation}
\frac{\sqrt{v^{2}\Delta^{2}+16w_{0}^{2}}+v\Delta}{3v}=\frac{s-1}{s}K,
\label{eq:speq}
\end{equation}
with
\begin{equation}
\Delta=\frac{s-1}{s}K+\Delta D.
\label{eq:Deltadef}
\end{equation}
For small $s-1$, we obtain
\begin{equation}
s-1=\frac{3w_{0}}{2v_{F}K},
\label{eq:s1}
\end{equation}
which gives $s=1.0183$, in agreement with our numerical result.

\section{STRAIN-INDUCED TOPOLOGICAL PHASE TRANSITION: QUANTITATIVE CHARACTERIZATION AND PROPOSED EXPERIMENTAL DETECTION}\label{sec:topology}

In this section we use our model to calculate the topological number and examine whether it changes when the energy gap closes.
\begin{figure}[htbp]
	\centering
	\includegraphics[width=\columnwidth]{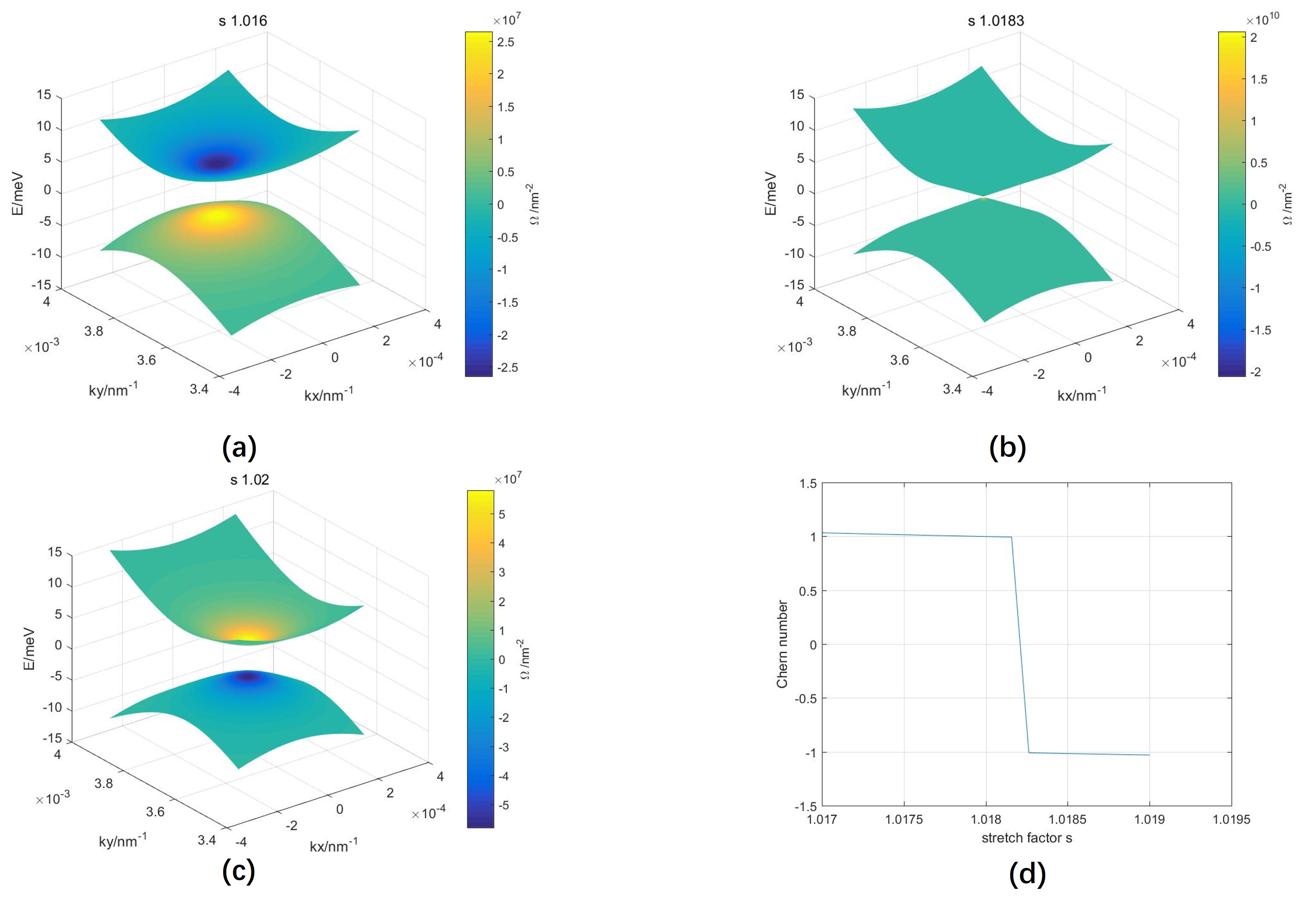}\\
	\caption{(a)--(c) Relationship between the Berry curvature and the energy band structure. We find a peak of the Berry curvature where the bands close, and, comparing (a) and (c), that the Berry curvature changes after the gap closing. (d) The change of the Chern number.}
	\label{Fig4}
\end{figure}

Since our model only describes the band structure near the Dirac point, it cannot capture the entire Brillouin zone. We may therefore not obtain a precise Chern number, but we can first calculate the Berry curvature and see whether it changes when the gap closes.

We performed the calculation using the efficient method of Fukui, Hatsugai, and Suzuki~\cite{Fukui2005}. The results in Fig.~\ref{Fig4} clearly show how the Berry curvature changes during the stretching. We find a topological phase transition when the gap closes: the Chern number changes from $1$ to $-1$, consistent with the DFT result~\cite{Su2025}.

We expect that this topological phase transition can be observed through the nonlinear Hall effect. Indeed, the nonlinear Hall effect has been studied in twisted double bilayer graphene~\cite{Chakraborty2022} and in strained Bernal-stacked bilayer graphene~\cite{He2026}, and the Berry-curvature dipole has been predicted to be strongly enhanced at topological phase transitions~\cite{Facio2018}. Since the SBLG system breaks inversion symmetry, the nonlinear Hall effect is allowed, and we can calculate the Berry-curvature dipole (BCD) density. Several experiments show that when the Berry curvature changes sign, the BCD also changes sign~\cite{Sinha2022,Li2022,Zhao2025}, so the direction of the Hall current reverses at the topological phase transition. Since we find a topological phase transition at a special strain factor, we expect the Hall current to reverse direction there.
\begin{figure}[htbp]
	\centering
	\includegraphics[width=\columnwidth]{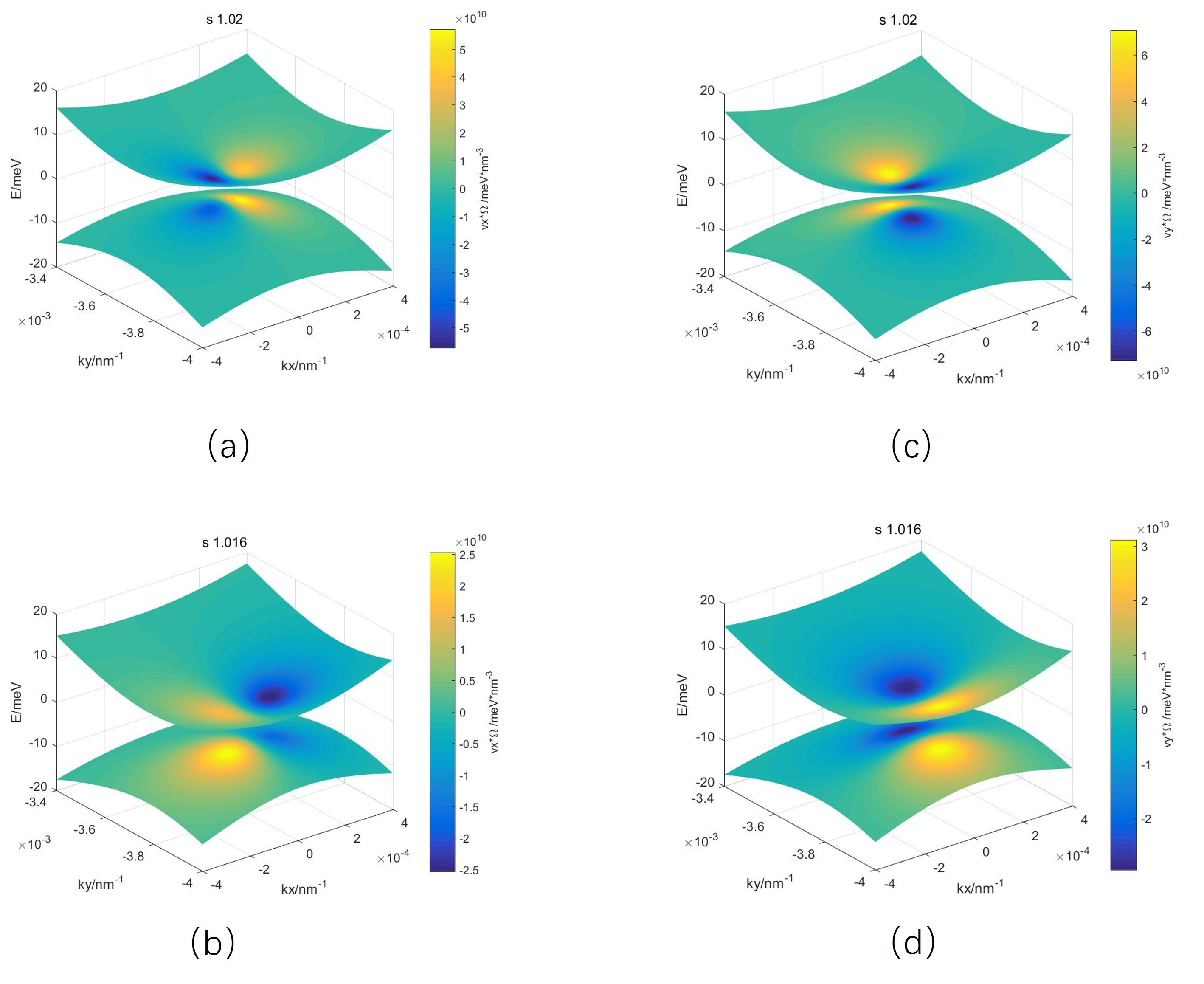}\\
        \caption{(a), (b) BCD density along the $k_{x}$ direction before and after the special strain factor. (c), (d) BCD density along the $k_{y}$ direction before and after the special strain factor. The Fermi level is $0$~meV. We calculate the BCD in this small region for each band.}
	\label{Fig5}
\end{figure}
We then calculate the BCD for both the conduction and valence bands in a small region near the degeneracy points~\cite{Sodemann2015}. For (a) $D_{\mathrm{conduction},x}=9.0\times10^{4}$~meV$^{-1}$nm$^{-1}$ and $D_{\mathrm{valence},x}=-9.0\times10^{4}$~meV$^{-1}$nm$^{-1}$; for (b) $D_{\mathrm{conduction},x}=-1.3\times10^{5}$~meV$^{-1}$nm$^{-1}$ and $D_{\mathrm{valence},x}=-1.3\times10^{5}$~meV$^{-1}$nm$^{-1}$; for (c) $D_{\mathrm{conduction},y}=6.5\times10^{3}$~meV$^{-1}$nm$^{-1}$ and $D_{\mathrm{valence},y}=6.5\times10^{3}$~meV$^{-1}$nm$^{-1}$; for (d) $D_{\mathrm{conduction},y}=-1.7\times10^{4}$~meV$^{-1}$nm$^{-1}$ and $D_{\mathrm{valence},y}=-1.7\times10^{4}$~meV$^{-1}$nm$^{-1}$. We find that the BCD changes sign at the topological phase transition. Along the $k_{y}$ direction, $D_{\mathrm{conduction},y}=D_{\mathrm{valence},y}$; since the two bands correspond to electrons and holes, the current involves $D_{\mathrm{conduction}}-D_{\mathrm{valence}}$, so no Hall current is produced along the $k_{y}$ direction. In real space, this means the Hall current cannot be observed along the axis perpendicular to the stretching direction, which is reasonable because the stretching only breaks inversion symmetry along the stretching direction.

\section{Conclusion}\label{sec:conclusion}
Conventional Bistritzer–MacDonald continuum models exclusively apply to twist-induced two-dimensional moir\'e superlattices and fail to capture the anisotropic one-dimensional moir\'e patterns originating from uniaxially strained bilayer graphene (SBLG). To fill this theoretical gap, we develop an original continuum model tailored for 1D moir\'e superlattices formed by single-direction stretching of AB-stacked bilayer graphene. In our formalism, interlayer hybridization is physically reinterpreted as inter-valley hopping between multiple Dirac cones, enabling an analytical explanation for the band gap closure at the Fermi level under a critical stretch factor $s\approx1.0183$, a phenomenon previously only observed numerically via TB and DFT. We further quantify the topological phase transition by evaluating Berry curvature and Chern numbers across the critical strain threshold, identifying an abrupt reversal of Chern number from $C=1$ to $C=-1$ upon gap closing. Compared with brute-force TB and DFT simulations, our analytical framework features concise mathematical forms and transparent physical intuition. This work establishes a universal low-energy theoretical paradigm for strain-engineered 1D moir\'e heterostructures and complements existing theories for twisted moir\'e systems, laying a foundational analytical tool for manipulating topological electronic states via uniaxial strain.

\begin{acknowledgments}

TL and JW would like to thank L. Su and L. Huang for their helpful discussion. JW was partially supported by the Natural Science Foundation of Guangdong Province (Grant Nos. 2017B030308003, 2019B121203002) and the Science, Technology and Innovation Commission of Shenzhen Municipality (Grant Nos. KYTDPT20181011104202253, JCYJ20170412152620376).
XRW acknowledges the support from  the Guangdong Provincial Quantum Strategy Special Project, the University Development Fund of the Chinese University of Hong Kong, Shenzhen, and 
the National Natural Science Foundation of China (Grants No. 12374122).
\par

The authors acknowledge the use of the AI-based writing tools Doubao and DeepSeek v4 pro for language polishing. The authors take full responsibility for the content of this manuscript.
\par

\emph{Data availability.}$-$The raw data supporting the findings of this manuscript and the custom codes for calculating data are available from the corresponding authors upon reasonable requests. 

\end{acknowledgments}

\bibliography{refs}
\onecolumngrid
\section*{Supplemental Material}

\section{Analysis of the $M$ matrix}
\label{sec:M}

In this section we discuss the degeneracy points for the different cases of the matrix $M$:
\begin{equation}
M=
\begin{pmatrix}
L\bigl[(a^2+(b+c)^2)E+2a(b+c)X\bigr]
& L\bigl[2abE+a^2\gamma+(b^2-c^2)\gamma^\dagger\bigr]+\gamma\\[6pt]
L\bigl[2abE+a^2\gamma^\dagger+(b^2-c^2)\gamma\bigr]+\gamma^\dagger
& L\bigl[(a^2+(b-c)^2)E+2a(b-c)X\bigr]
\end{pmatrix},
\label{eq:M}
\end{equation}
with
\begin{equation}
L=\frac{1}{E^2-\gamma^\dagger\gamma}.
\label{eq:L}
\end{equation}
In fact, we do not need to solve the equation. For a $2\times2$ Hamiltonian, a degeneracy point appears when $M_{11}=M_{22}=E$ and $\operatorname{Im}M_{12}=\operatorname{Re}M_{12}=0$. If we can solve these equations, we can obtain the information about the degeneracy points. We divide the problem into different situations. We first discuss the case in which the two $K$ valleys are at the same position ($\Delta=0$) in Sec.~\ref{sec:delta0}, and then the case in which they are displaced ($\Delta\neq0$) in Sec.~\ref{sec:deltaneq0}.

\subsection{Two $K$ valleys at the same position ($\Delta=0$)}
\label{sec:delta0}

\subsubsection{The case $b=c=0$}
\label{sec:bc0}

\begin{figure}[t]
\includegraphics[width=\columnwidth]{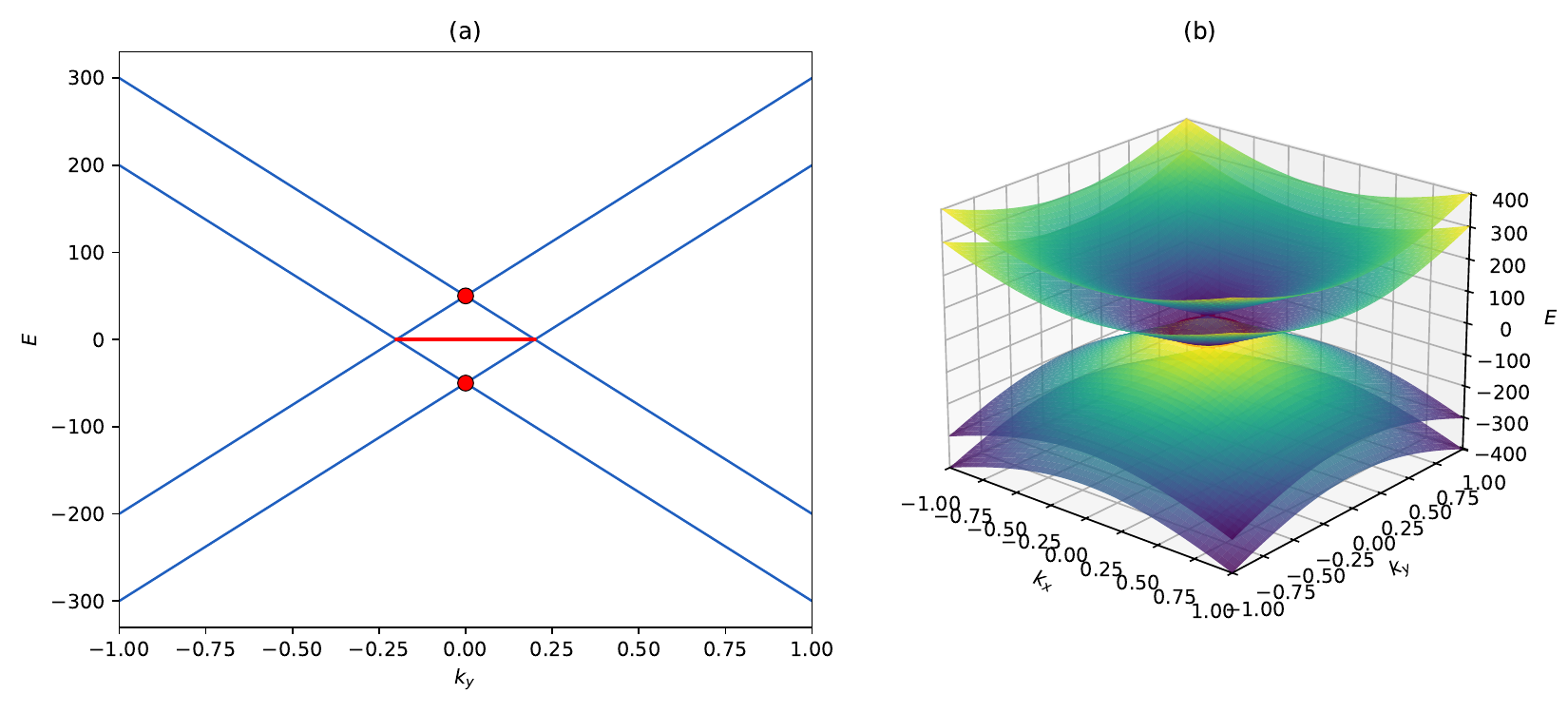}
\caption{(a) Band structure in the $k_x=0$ cross section: the degeneracy circle projects onto the segment at $E=0$ and the two points appear at $E=\pm a$. (b) Full energy band structure.}
\label{fig:1}
\end{figure}

If $b$ and $c$ are zero, the $M$ matrix becomes
\begin{equation}
M=
\begin{pmatrix}
LEa^2 & (LEa^2+1)\gamma\\
(LEa^2+1)\gamma^\dagger & LEa^2
\end{pmatrix}.
\label{eq:Mbc0}
\end{equation}
By setting $M_{11}=M_{22}=E$ and $\operatorname{Im}M_{12}=\operatorname{Re}M_{12}=0$, we obtain the location of the degeneracy points:
\begin{align}
\gamma&=0,\qquad E=\pm a,\label{eq:gamma0}\\
\gamma^\dagger\gamma&=a^2,\qquad E=0.\label{eq:circle}
\end{align}
The locations given by Eqs.~\eqref{eq:gamma0} and \eqref{eq:circle} are shown in Fig.~\ref{fig:1}(a). The energy band structure of the $4\times4$ matrix with $b=c=0$ is shown in Fig.~\ref{fig:1}(b), from which we can confirm that our result is correct.

\subsubsection{The case $c=0$}
\label{sec:c0}

\begin{figure}[t]
\includegraphics[width=\columnwidth]{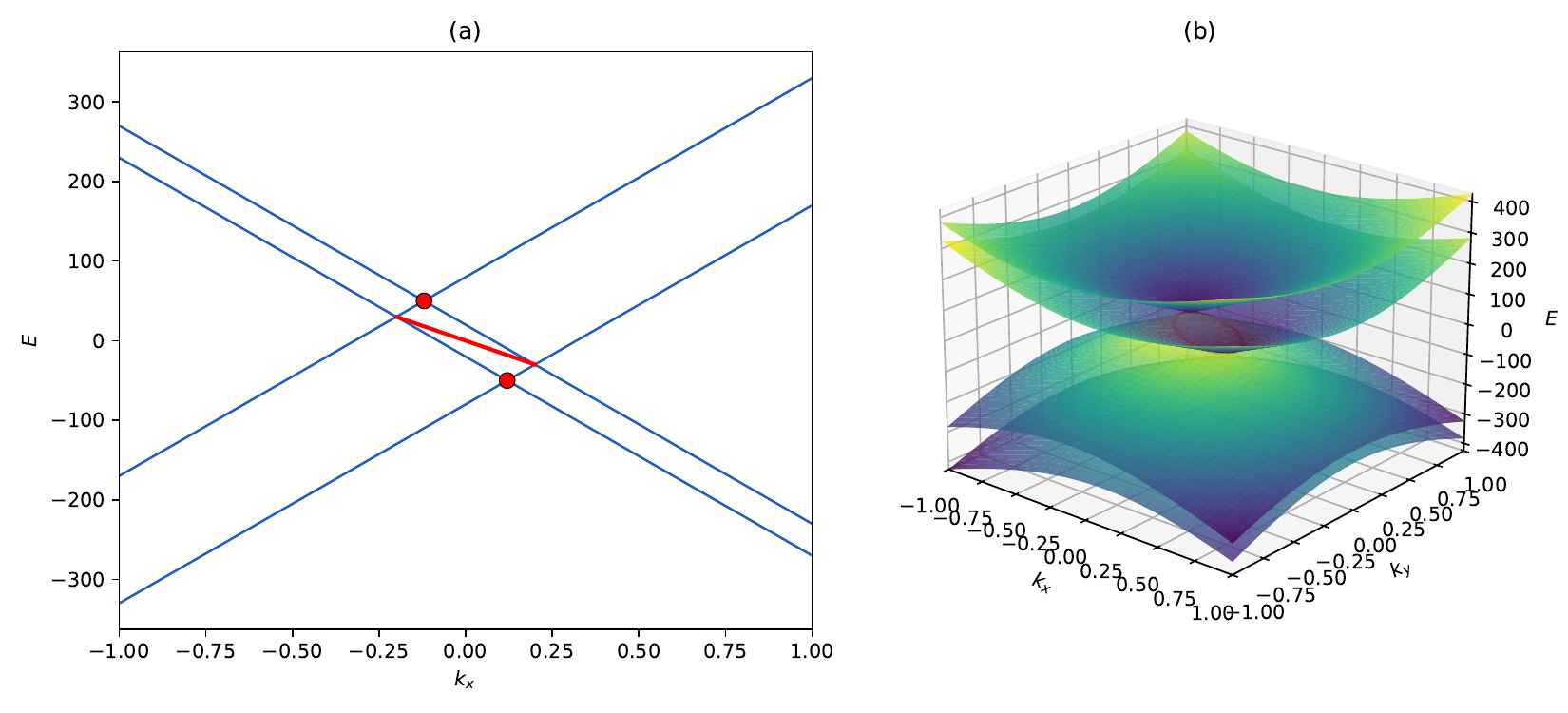}
\includegraphics[width=\columnwidth]{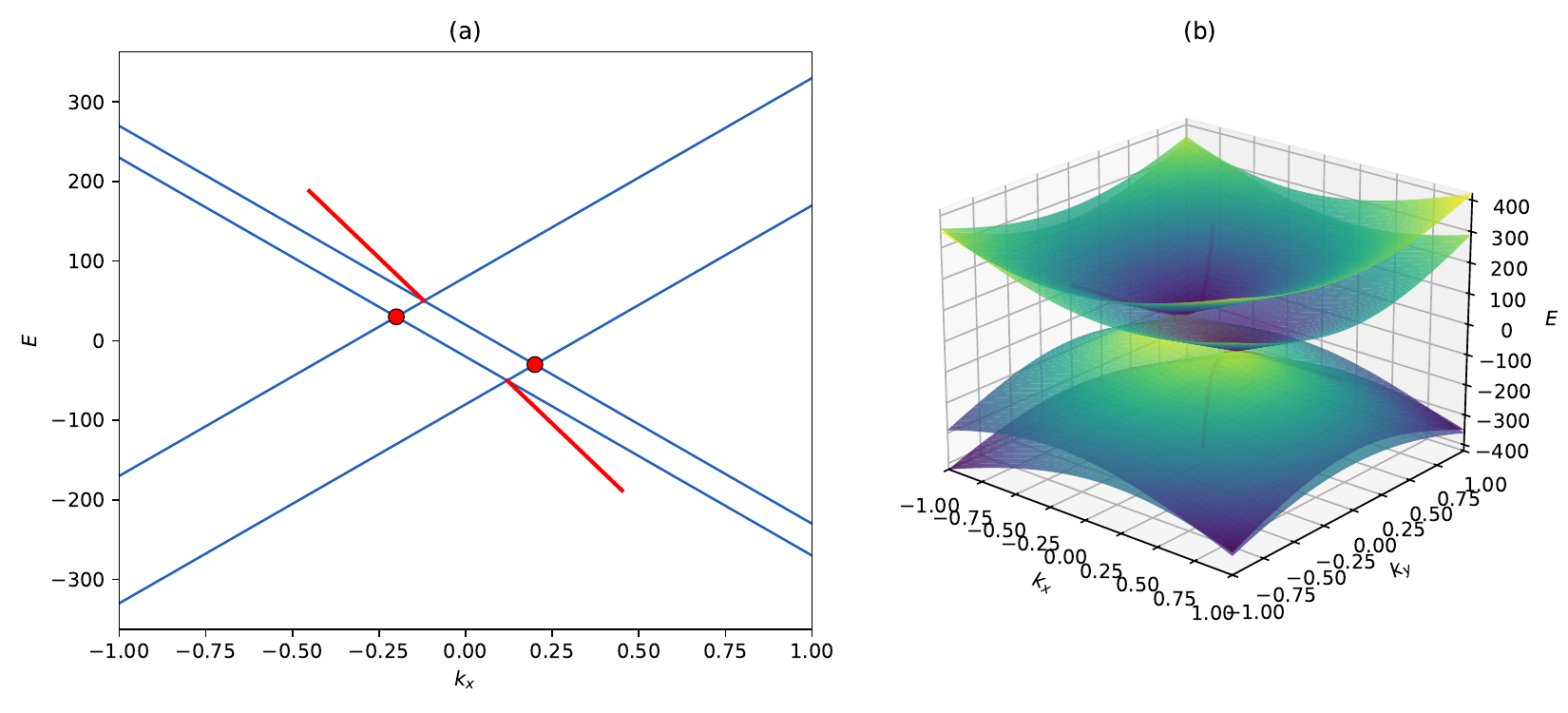}
\caption{The two different situations for $c=0$: (a) $a^2>b^2$, an ellipse; (b) $a^2<b^2$, a hyperbola. Each panel shows the band structure in the $k_y=0$ cross section together with the projected degeneracy curve and points.}
\label{fig:2}
\end{figure}

Next, we set only $c$ to zero. The $M$ matrix becomes
\begin{equation}
M=
\begin{pmatrix}
L\bigl[E(a^2+b^2)+2abX\bigr]
& L\bigl[2abE+a^2\gamma+b^2\gamma^\dagger\bigr]+\gamma\\[6pt]
L\bigl[2abE+a^2\gamma^\dagger+b^2\gamma\bigr]+\gamma^\dagger
& L\bigl[E(a^2+b^2)+2abX\bigr]
\end{pmatrix}.
\label{eq:Mc0}
\end{equation}
This calculation is a bit more involved, so we write it down here. The equations are
\begin{align}
E&=L\bigl[E(a^2+b^2)+2abX\bigr],\label{eq:c0a}\\
2abLE+\bigl[(a^2+b^2)L+1\bigr]X&=0,\label{eq:c0b}\\
\bigl[(-a^2+b^2)L-1\bigr]Y&=0.\label{eq:c0c}
\end{align}
For Eq.~\eqref{eq:c0c} we can first set $Y=0$. Then Eqs.~\eqref{eq:c0a} and \eqref{eq:c0b} become
\begin{align}
-X^3+XE^2+2abE+X(a^2+b^2)&=0,\label{eq:c0d}\\
-E^3+EX^2+2abX+E(a^2+b^2)&=0.\label{eq:c0e}
\end{align}
Using $a^3+b^3=(a+b)(a^2-ab+b^2)$ and $a^3-b^3=(a-b)(a^2+ab+b^2)$ and adding Eqs.~\eqref{eq:c0d} and \eqref{eq:c0e}, we obtain
\begin{equation}
(E-X)^2=(a+b)^2,\qquad (E+X)^2=(a-b)^2.
\label{eq:c0f}
\end{equation}
Thus the solutions of $(X,E)$ are $(-b,a)$, $(b,-a)$, $(-a,b)$, and $(a,-b)$. If $Y$ is nonzero, Eq.~\eqref{eq:c0c} gives
\begin{equation}
L=\frac{1}{b^2-a^2},\qquad\text{i.e.,}\quad E^2-X^2-Y^2=b^2-a^2.
\label{eq:c0g}
\end{equation}
Equations~\eqref{eq:c0a} and \eqref{eq:c0b} are then identical and give $X=-(a/b)E$. Together we obtain the equation
\begin{equation}
a^2(a^2-b^2)=(a^2-b^2)X^2+a^2Y^2.
\label{eq:conic}
\end{equation}
Clearly this is a conic section: when $a^2>b^2$ it is an ellipse, and when $a^2<b^2$ it is a hyperbola. The solution points $(X,E)=(\pm a,\mp b)$ also lie on the curve. Figure~\ref{fig:2} shows the two situations.

\subsubsection{The case $b=0$}
\label{sec:b0}

\begin{figure}[t]
\includegraphics[width=\columnwidth]{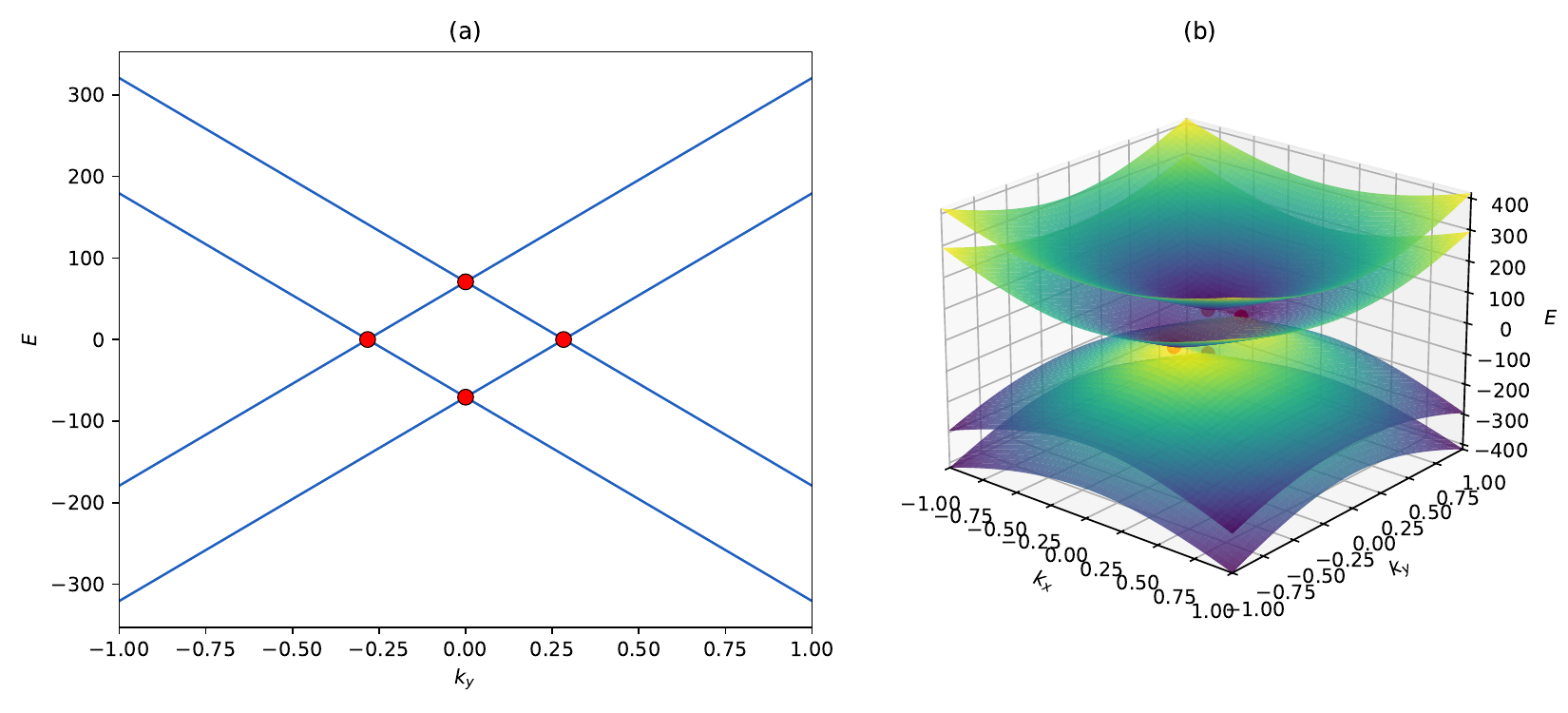}
\caption{(a) Band structure in the $k_x=0$ cross section with the four degeneracy points. (b) Full energy band structure.}
\label{fig:3}
\end{figure}

Next, we set only $b$ to zero. The $M$ matrix becomes
\begin{equation}
M=
\begin{pmatrix}
L\bigl[E(a^2+c^2)+2acX\bigr]
& L\bigl[a^2\gamma-c^2\gamma^\dagger\bigr]+\gamma\\[6pt]
L\bigl[a^2\gamma^\dagger-c^2\gamma\bigr]+\gamma^\dagger
& L\bigl[E(a^2+c^2)-2acX\bigr]
\end{pmatrix}.
\label{eq:Mb0}
\end{equation}
The equations are
\begin{align}
L\bigl[E(a^2+c^2)+2acX\bigr]&=L\bigl[E(a^2+c^2)-2acX\bigr]=E,\label{eq:b0a}\\
\bigl[L(a^2-c^2)+1\bigr]X&=0,\label{eq:b0b}\\
\bigl[L(-a^2-c^2)-1\bigr]Y&=0.\label{eq:b0c}
\end{align}
The solutions to this group of equations are four points, of the form $(X,Y,E)$:
\begin{equation}
(0,0,\pm\sqrt{a^2+c^2}),\qquad (0,\pm\sqrt{a^2+c^2},0).
\label{eq:b0sol}
\end{equation}
This means there are four degeneracy points, as shown in Fig.~\ref{fig:3}(a).

\subsubsection{The case where $a$, $b$, and $c$ are all nonzero}
\label{sec:allnz0}

\begin{figure}[t]
\includegraphics[width=\columnwidth]{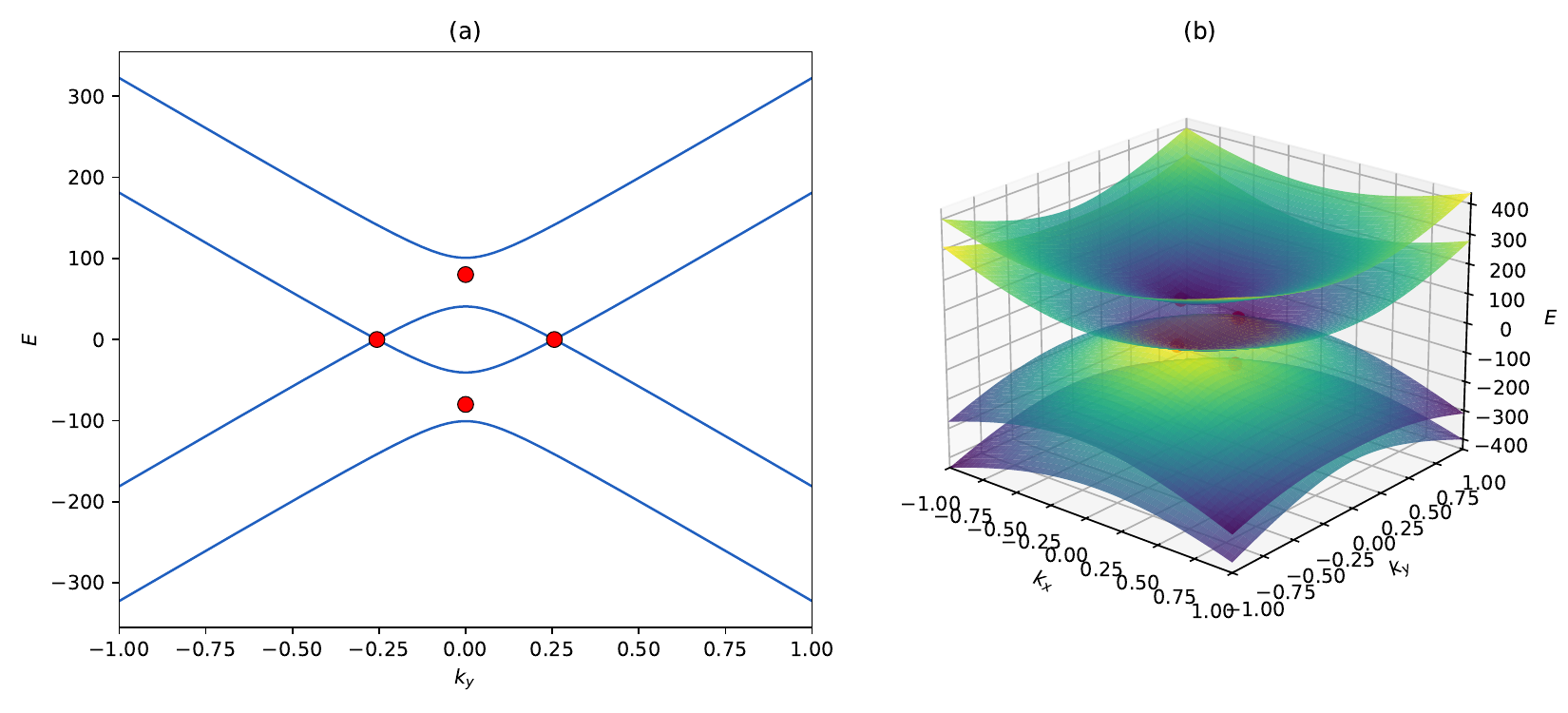}
\caption{Band structure for $a$, $b$, $c$ all nonzero: (a) $k_x=0$ cross section with the four degeneracy points; (b) full band structure.}
\label{fig:4}
\end{figure}

Finally, if $a$, $b$, and $c$ are all nonzero real numbers, the $M$ matrix is given by Eq.~\eqref{eq:M}. Thus we obtain the equations
\begin{align}
L\bigl[(a^2+(b+c)^2)E+2a(b+c)X\bigr]&=L\bigl[(a^2+(b-c)^2)E+2a(b-c)X\bigr]=E,\label{eq:alla}\\
2LEab+\bigl(a^2L+(b+c)(b-c)L+1\bigr)X&=0,\label{eq:allb}\\
\bigl(-a^2L+(b+c)(b-c)L-1\bigr)Y&=0.\label{eq:allc}
\end{align}
The solutions of this group of equations are also four points, roughly distributed as in Fig.~\ref{fig:4}(a):
\begin{align}
E&=0,\qquad X=0,\qquad Y=\pm\sqrt{a^2-(b+c)(b-c)},\label{eq:allsol1}\\
E&=\pm\sqrt{\frac{4a^2\bigl(a^2-(b+c)(b-c)\bigr)}{4a^2-4b^2}},\qquad X=-\frac{b}{a}E,\qquad Y=0.\label{eq:allsol2}
\end{align}

\subsection{Two $K$ valleys displaced along the $k_x$ axis ($\Delta\neq0$)}
\label{sec:deltaneq0}

Now we discuss a more complex situation, in which the two $K$ valleys are not at the same position. As in the stretched-bilayer-graphene (SBLG) system, we displace one $K$ valley along the $k_x$ axis by a momentum $\bm{\Delta}=(\Delta,0)$. We label the two $K$ valleys as $\alpha$ and $\beta$. The total Hamiltonian is written as
\begin{equation}
H=
\begin{pmatrix}
h(k)&T\\
T^\dagger&h(k+\Delta)
\end{pmatrix}
=
\begin{pmatrix}
h_\alpha&T\\
T^\dagger&h_\beta
\end{pmatrix},
\label{eq:Hdisp}
\end{equation}
where
\begin{equation}
h_\alpha=
\begin{pmatrix}
0&\gamma_\alpha\\
\gamma_\alpha^\dagger&0
\end{pmatrix}
=
\begin{pmatrix}
0&X-iY\\
X+iY&0
\end{pmatrix},
\label{eq:ha}
\end{equation}
\begin{equation}
h_\beta=
\begin{pmatrix}
0&\gamma_\beta\\
\gamma_\beta^\dagger&0
\end{pmatrix}
=
\begin{pmatrix}
0&X+v\Delta-iY\\
X+v\Delta+iY&0
\end{pmatrix}.
\label{eq:hb}
\end{equation}
The $M$ matrix becomes
\begin{equation}
M=
\begin{pmatrix}
L_\beta\bigl[(a^2+(b+c)^2)E+2a(b+c)X_\beta\bigr]
& L_\beta\bigl[2abE+a^2\gamma_\beta+(b^2-c^2)\gamma_\beta^\dagger\bigr]+\gamma_\alpha\\[6pt]
L_\beta\bigl[2abE+a^2\gamma_\beta^\dagger+(b^2-c^2)\gamma_\beta\bigr]+\gamma_\alpha^\dagger
& L_\beta\bigl[(a^2+(b-c)^2)E+2a(b-c)X_\beta\bigr]
\end{pmatrix},
\label{eq:Mdisp}
\end{equation}
where
\begin{equation}
L_\beta=\frac{1}{E^2-\gamma_\beta^\dagger\gamma_\beta},\qquad X_\beta=X+v\Delta.
\label{eq:Lbeta}
\end{equation}
As in the last subsection, we again divide the problem into different situations.

\subsubsection{The case $b=c=0$}
\label{sec:bc0disp}

\begin{figure}[t]
\includegraphics[width=\columnwidth]{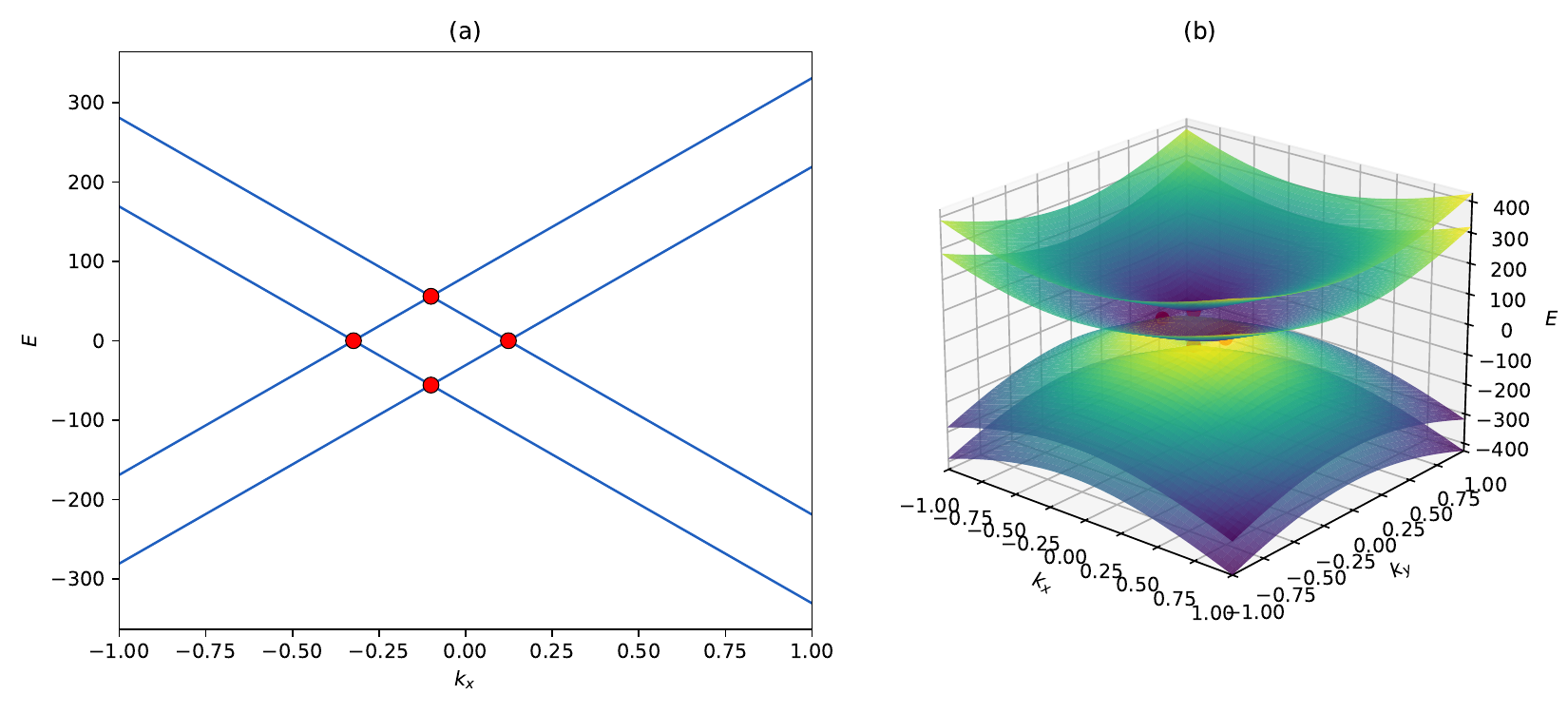}
\caption{Band structure for $b=c=0$ with $\Delta\neq0$: (a) $k_y=0$ cross section with the four degeneracy points; (b) full band structure.}
\label{fig:5}
\end{figure}

If $b$ and $c$ are zero, the $M$ matrix becomes
\begin{equation}
M=
\begin{pmatrix}
L_\beta Ea^2 & L_\beta Ea^2\gamma_\beta+\gamma_\alpha\\
L_\beta Ea^2\gamma_\beta^\dagger+\gamma_\alpha^\dagger & L_\beta Ea^2
\end{pmatrix}.
\label{eq:Mdisp0}
\end{equation}
The equations are
\begin{align}
L_\beta a^2\gamma_\beta+\gamma_\alpha&=0,\label{eq:disp0a}\\
L_\beta a^2E&=E.\label{eq:disp0b}
\end{align}
The results are four points:
\begin{align}
X&=\frac{-v\Delta\pm\sqrt{v^2\Delta^2+4a^2}}{2},\qquad Y=0,\qquad E=0,\label{eq:disp0c}\\
X&=-\frac{v\Delta}{2},\qquad Y=0,\qquad E=\pm\sqrt{a^2+\frac{v^2\Delta^2}{4}}.\label{eq:disp0d}
\end{align}
Figure~\ref{fig:5} shows the energy band structure.

\subsubsection{The case $c=0$}
\label{sec:c0disp}

\begin{figure}[t]
\includegraphics[width=\columnwidth]{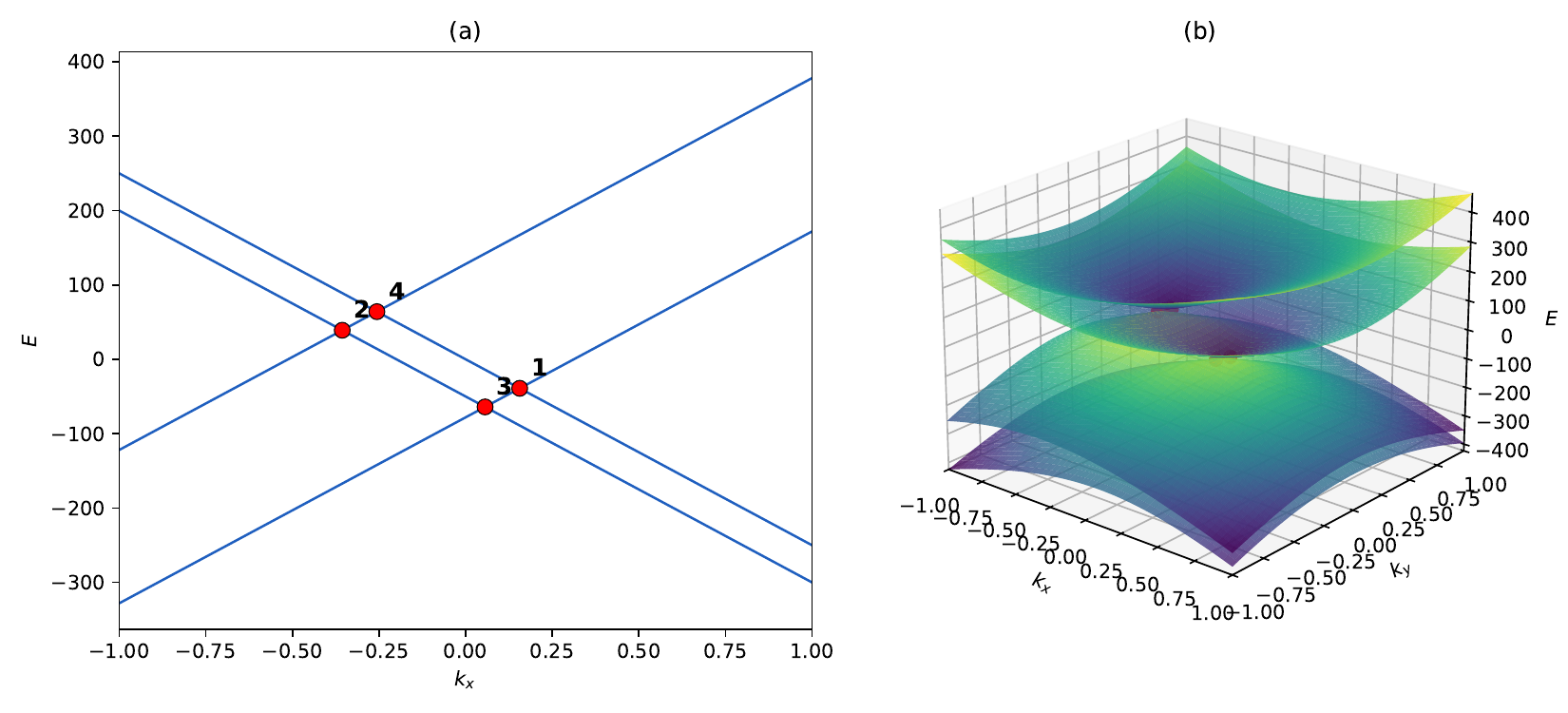}
\caption{Band structure for $c=0$ with $\Delta\neq0$: (a) $k_y=0$ cross section with the four degeneracy points labeled $1$--$4$; (b) full band structure.}
\label{fig:6}
\end{figure}

Next, if $c$ is zero, the $M$ matrix becomes
\begin{equation}
M=
\begin{pmatrix}
L_\beta\bigl[(a^2+b^2)E+2abX_\beta\bigr]
& L_\beta\bigl[2abE+a^2\gamma_\beta+b^2\gamma_\beta^\dagger\bigr]+\gamma_\alpha\\[6pt]
L_\beta\bigl[2abE+a^2\gamma_\beta^\dagger+b^2\gamma_\beta\bigr]+\gamma_\alpha^\dagger
& L_\beta\bigl[(a^2+b^2)E+2abX_\beta\bigr]
\end{pmatrix}.
\label{eq:Mc0disp}
\end{equation}
The equations are
\begin{align}
L_\beta\bigl[(a^2+b^2)E+2abX_\beta\bigr]&=E,\label{eq:c0dispa}\\
L_\beta\bigl[2abE+a^2X_\beta+b^2X_\beta\bigr]+X_\alpha&=0,\label{eq:c0dispb}\\
L_\beta\bigl[a^2Y_\beta-b^2Y_\beta\bigr]+Y_\alpha&=0.\label{eq:c0dispc}
\end{align}
Defining $R_1=\sqrt{v^2\Delta^2+4(a+b)^2}$ and $R_2=\sqrt{v^2\Delta^2+4(a-b)^2}$, the solutions are the four points
\begin{align}
X_1&=-\frac{v\Delta}{2}+\frac{R_1+R_2}{4},&\qquad E_1&=-\frac{R_1-R_2}{4},\\
X_2&=-\frac{v\Delta}{2}-\frac{R_1+R_2}{4},&\qquad E_2&=+\frac{R_1-R_2}{4},\\
X_3&=-\frac{v\Delta}{2}+\frac{R_1-R_2}{4},&\qquad E_3&=-\frac{R_1+R_2}{4},\\
X_4&=-\frac{v\Delta}{2}-\frac{R_1-R_2}{4},&\qquad E_4&=+\frac{R_1+R_2}{4}.
\end{align}
Figure~\ref{fig:6} shows the energy band structure. From Fig.~\ref{fig:6}(a) we can see that this type of $T$ hopping displaces the two $K$ valleys along the $k_x$ axis: the $\alpha$ valley moves to point 2 and the $\beta$ valley moves to point 3.

\subsubsection{The case $b=0$}
\label{sec:b0disp}

Next, we set only $b$ to zero. The $M$ matrix becomes
\begin{equation}
M=
\begin{pmatrix}
L_\beta\bigl[(a^2+c^2)E+2acX_\beta\bigr]
& L_\beta\bigl[a^2\gamma_\beta-c^2\gamma_\beta^\dagger\bigr]+\gamma_\alpha\\[6pt]
L_\beta\bigl[a^2\gamma_\beta^\dagger-c^2\gamma_\beta\bigr]+\gamma_\alpha^\dagger
& L_\beta\bigl[(a^2+c^2)E-2acX_\beta\bigr]
\end{pmatrix}.
\label{eq:Mb0disp}
\end{equation}
The equations are
\begin{align}
L_\beta\bigl[(a^2+c^2)E+2acX_\beta\bigr]&=L_\beta\bigl[(a^2+c^2)E-2acX_\beta\bigr]=E,\label{eq:b0dispa}\\
L_\beta\bigl[a^2X_\beta-c^2X_\beta\bigr]+X_\alpha&=0,\label{eq:b0dispb}\\
L_\beta\bigl[a^2Y_\beta+c^2Y_\beta\bigr]+Y_\alpha&=0.\label{eq:b0dispc}
\end{align}
Clearly Eq.~\eqref{eq:b0dispa} gives $X_\beta=0$, and then Eq.~\eqref{eq:b0dispb} gives $X_\alpha=0$. This contradicts our initial condition $X_\beta=X_\alpha+v\Delta$. Thus the equations have no solution, which means that in this case there is no degeneracy point.

\subsubsection{The case where $a$, $b$, and $c$ are all nonzero}
\label{sec:allnzd}

The case where $a$, $b$, and $c$ are all nonzero is the same as the case $b=0$ above, and there is no degeneracy point.

\section{Interlayer interaction term}
\label{sec:interlayer}

In real space, we need to focus on these four terms:
\begin{equation}
\begin{aligned}
& C_{\alpha,A}(r_{\alpha,A})\,t\bigl(|r_{\alpha,A}-r_{\beta,B}|\bigr)\,C_{\beta,B}(r_{\beta,B})^\dagger\\
& C_{\alpha,A}(r_{\alpha,A})\,t\bigl(|r_{\alpha,A}-r_{\beta,C}|\bigr)\,C_{\beta,C}(r_{\beta,C})^\dagger\\
& C_{\alpha,B}(r_{\alpha,B})\,t\bigl(|r_{\alpha,B}-r_{\beta,C}|\bigr)\,C_{\beta,C}(r_{\beta,C})^\dagger\\
& C_{\alpha,B}(r_{\alpha,B})\,t\bigl(|r_{\alpha,B}-r_{\beta,D}|\bigr)\,C_{\beta,D}(r_{\beta,D})^\dagger
\end{aligned}
\label{eq:inter1}
\end{equation}

\begin{figure}[t]
\includegraphics[width=\columnwidth]{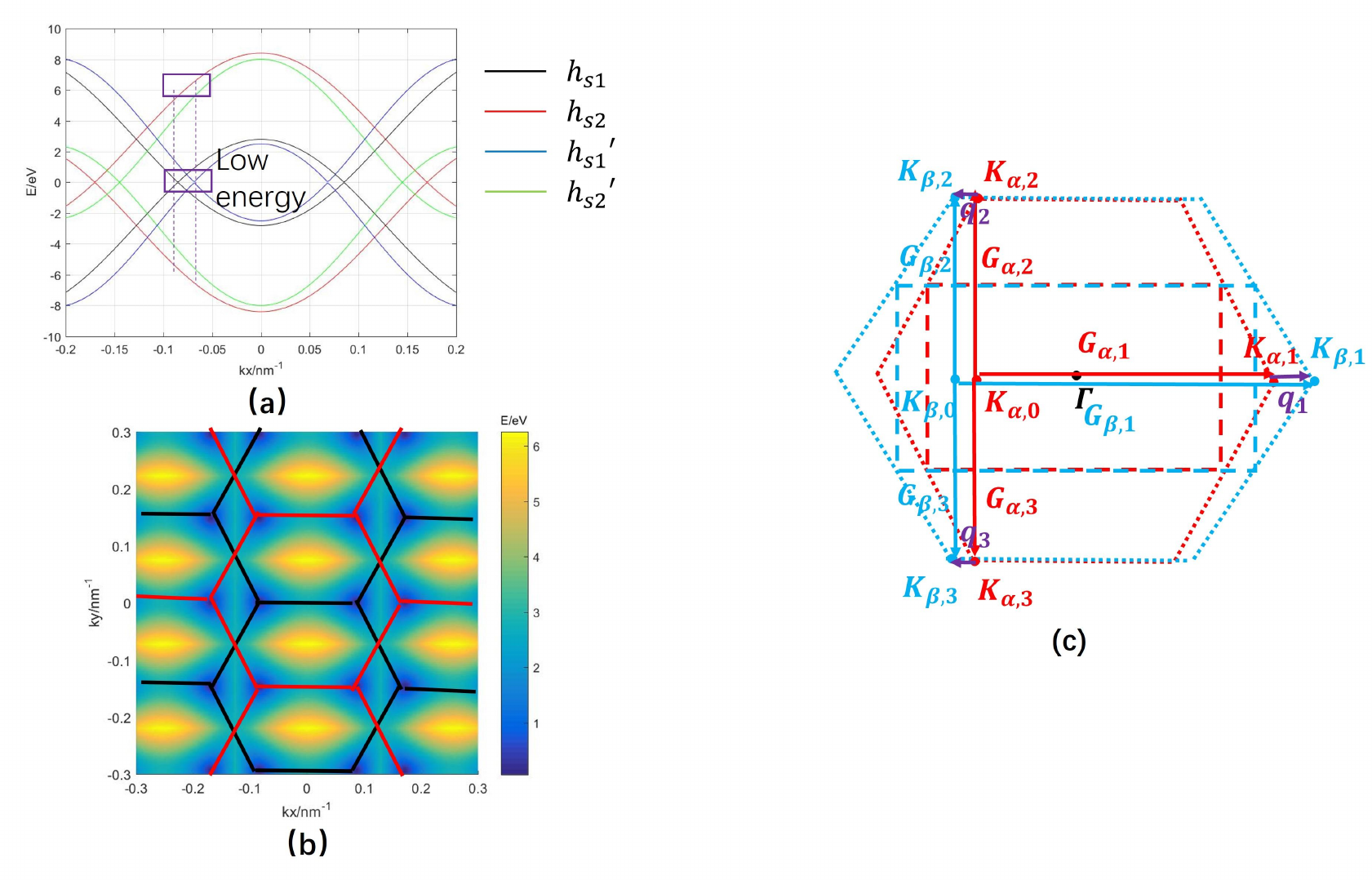}
\caption{(a) Band structure of the stretched-bilayer system without the interlayer interaction. The eight bands are given by the four Hamiltonians $h_{si}$ and $h'_{si}$ after the change of basis. In our study we only consider the low-energy part; when we study the system near the Dirac cone, we need only keep the interaction between $h_{s1}$ and $h'_{s1}$, because the other two bands have larger energy. (b) Honeycomb structure of $h_{s1}$ (black) and $h'_{s1}$ (red). (c) The red dashed rectangle is the first Brillouin zone of the $\alpha$ layer and the blue dashed rectangle is the first Brillouin zone of the $\beta$ layer. By applying the three groups of reciprocal lattice vectors $G_{l,i}$, we find that the three types of interaction can be regarded as the hopping between three groups of $K_{l,i}$. For the three $K$ valleys from the same layer, we find that they belong to the same honeycomb structure $h_{s1}$ or $h'_{s1}$, shown by the dotted lines. This also proves the conclusion that the hopping between $h_{s1}$ and $h'_{s1}$ is independent.}
\label{fig:7}
\end{figure}

Using the Fourier transformation, we can rewrite the expression in reciprocal space as
\begin{equation}
H(k_\alpha,k_\beta)_{\alpha,\beta}=
\sum_{k_\alpha,k_\beta} C_{\alpha,i}(k_\alpha)^\dagger C_{\beta,j}(k_\beta)
\sum_{R_{\alpha,i},R_{\beta,j},q}
e^{-i(k_\alpha-q)R_{\alpha,i}}\,t(q)\,e^{i(k_\beta-q)R_{\beta,j}}.
\label{eq:inter2}
\end{equation}
Here $R_{l,i}$ is the position of atom $i$ in layer $l$. It can be decomposed into two parts, $R_{l,i}=R_l+\tau_{l,i}$, where $R_l$ describes the position of the unit cell and $\tau_{l,i}$ describes the position of the atom within the unit cell. For an SBLG system that starts in AB stacking, the values of $\tau_{l,i}$ are
\begin{equation}
\tau_{\alpha,A}=(0,d),\qquad \tau_{\alpha,B}=(0,0),\qquad \tau_{\beta,B}=(0,-d),\qquad
\tau_{\beta,C}=\Bigl(-\tfrac{\sqrt{3}\,s}{2}d,-\tfrac{3}{2}d\Bigr),\qquad
\tau_{\beta,D}=\Bigl(-\tfrac{\sqrt{3}\,s}{2}d,\tfrac{1}{2}d\Bigr).
\label{eq:tau}
\end{equation}
Using the sum rule $\sum_{R_l}e^{ikR_l}=\sum_{G_l}\delta_{-G_l,k}$, where $G_l$ is the reciprocal lattice vector of layer $l$, we can rewrite the second sum as
\begin{equation}
\sum_{G_\alpha,G_\beta,q,\tau_{\alpha,i},\tau_{\beta,j}}
\delta_{k_\alpha-q,-G_\alpha}\,\delta_{k_\beta-q,-G_\beta}\,
e^{-i(k_\alpha-q)\tau_{\alpha,i}}\,t(q)\,e^{i(k_\beta-q)\tau_{\beta,j}}.
\label{eq:inter3}
\end{equation}
We expand the Hamiltonian near $K_{\alpha,0}$ and $K_{\beta,0}$. This gives $k_l=K_{l,0}+q_l$, where $K_{l,0}$ is the momentum of the $K$ valley and $q_l$ is much smaller than $K_{l,0}$. The first $\delta$ function gives $q=K_{\alpha,0}+q_\alpha+G_\alpha$. Thus we can rewrite the sum as
\begin{equation}
\sum_{G_\alpha,G_\beta,\tau_\alpha,\tau_\beta}
\delta_{K_{\beta,0}+q_\beta+G_\beta,\,K_{\alpha,0}+q_\alpha+G_\alpha}\,
e^{iG_\alpha\tau_\alpha}\,t(K_{\alpha,0}+q_\alpha+G_\alpha)\,
e^{i(K_{\beta,0}-K_{\alpha,0}+q_\beta-q_\alpha-G_\alpha)\tau_\beta}.
\label{eq:inter4}
\end{equation}
We can always find a $K$ valley $K'_l$ such that $K'_l=K_l+G_l$. Thus $t(K_\alpha+q_\alpha+G_\alpha)\approx t(K'_\alpha)$. The expression of the interlayer interaction factor $t$ is
\begin{equation}
t(q)=t_0\exp\bigl[-\alpha(q\,d_{\alpha\beta})^\gamma\bigr]=t_0\exp\bigl[-\alpha\bigl(|q|\,|d_{\alpha\beta}|\bigr)^\gamma\bigr].
\label{eq:inter5}
\end{equation}
Here $t_0=2\,\mathrm{eV}\,\text{\AA}^2$, $\alpha=0.13$, $\gamma=1.25$, and $d_{\alpha\beta}$ is the vertical distance between the two layers. Note that the vector $d_{\alpha\beta}$ is always perpendicular to the vector $q$. When we select the three reciprocal vectors $G_{\alpha,1}$, $G_{\alpha,2}$, and $G_{\alpha,3}$ shown in Fig.~\ref{fig:7}, $|q|=K$. For other $G_\alpha$, the value of $t(q)$ is much smaller than $t(K)$. Thus we get
\begin{equation}
q_\beta-q_\alpha=(K_{\beta,0}+G_{\beta,j})-(K_{\alpha,0}+G_{\alpha,i})=K_{\beta,j}-K_{\alpha,i}.
\label{eq:inter6}
\end{equation}
Since $q_\beta$ and $q_\alpha$ are much smaller than $K$, we find $i=j$; otherwise $|K_{\beta,j}-K_{\alpha,i}|$ would be larger than $K$. Thus Eq.~\eqref{eq:inter6} can be rewritten as
\begin{equation}
q_\beta-q_\alpha=K_{\beta,i}-K_{\alpha,i}=-q_i,
\label{eq:inter7}
\end{equation}
where $i=1,2,3$.

Thus we set a new variable $k=q_\alpha$, and the interlayer coupling term near the pair $K_{\alpha,0}$ and $K_{\beta,0}$ can be rewritten as
\begin{equation}
H(k)^{(1)}_{\alpha,\beta}=
\sum_{k,i} C_{\alpha,0}(K_{\alpha,0}+k)^\dagger\,t(K)\,
e^{i(G_{\alpha,i}\tau_{\alpha,m}-G_{\beta,i}\tau_{\beta,n})}\,
C_{\beta,j}(K_{\beta,0}+k-q_i).
\label{eq:inter8}
\end{equation}
Here $t(K)=w_0$ is a constant. We also find that $q_2=q_3$. This means that there are two types of hopping, $T_1$ and $T_2$. Each of them is a $2\times2$ matrix:
\begin{equation}
T_1=w_0\begin{pmatrix}1&1\\1&1\end{pmatrix},\qquad
T_2=w_0\begin{pmatrix}-1&2\\-1&-1\end{pmatrix}.
\label{eq:inter9}
\end{equation}

\section{Differences from previous Bistritzer--MacDonald-like continuum models of stretched graphene \& the periodic along $\mathbf{b_{1}^{supper}}$}
\label{sec:comparison}

In this section we compare our model with the model of Ref.~\cite{Ji2025} and gives the low-energy dispersion of $h_{sg}'$. Their description of the SBLG system is based on the twisted-bilayer-graphene model of Bistritzer and MacDonald~\cite{BistritzerMacDonald}, and it can describe stretching of the graphene layer in two directions. When focusing on stretching along the $x$ direction, the matrix obtained from their model is the same as the low-energy part of our matrix. However, their model does not contain the high-energy part $h_{sg}'$ given by Eq.~\eqref{eq:hSBLGblock}. On the other hand, our calculation shows that the first Brillouin zone of the superlattice is a rectangle spanned by $\mathbf{b}_{1}^{\mathrm{super}}=(0,\frac{\sqrt{3}}{2}K)$ and $\mathbf{b}_{2}^{\mathrm{super}}=(\frac{3}{2}\frac{s-1}{s}K,0)$, which means that the Hamiltonian in momentum space must be periodic in both the $x$ and $y$ directions, even though the layer is stretched only along the $x$ direction. In contrast, their model loses the periodicity along the $y$ direction when stretching only along $x$. We show below that this periodicity, which is absent in the low-energy sector, is actually encoded in the high-energy terms.

Focusing on the single-layer part, we have $h_{s1}(\mathbf{k}+\mathbf{b}_{1}^{\mathrm{super}})=h_{s2}(\mathbf{k})$ and $h_{s1}'(\mathbf{k}+\mathbf{b}_{1}^{\mathrm{super}})=h_{s2}'(\mathbf{k})$; that is, after advancing by one period along the $y$ direction, the originally high-energy part becomes the low-energy part (this can also be seen in Fig.~\ref{Fig2}). We can therefore still use the method of Sec.~\ref{sec:interlayer} to obtain the interaction term. The only difference is that now we calculate the term $T_{\alpha,\beta}'$, given by
\begin{equation}
T_{\alpha,\beta}'=\begin{bmatrix}
-w_{5} & w_{6}\\
-w_{7} & -w_{8}
\end{bmatrix}.
\label{eq:Tabp}
\end{equation}
Thus, in Eq.~\eqref{eq:inter8}, the position $\tau$ in real space should be taken as the point corresponding to $w_i$ in Table~\ref{Tab1}. We then obtain
\begin{equation}
T'_1=w_0\begin{pmatrix}-1&1\\-1&-1\end{pmatrix},\qquad
T'_2=w_0\begin{pmatrix}1&2\\1&1\end{pmatrix}.
\label{eq:T1T2p}
\end{equation}
Combining the single-layer relations above with the coupling matrices $T_i'$ of the high-energy sector, the periodicity of the full Hamiltonian is restored. Since $\mathbf{b}_{1}^{\mathrm{super}}=(0,\frac{\sqrt{3}}{2}K)$ is the reciprocal lattice vector $\mathbf{G}_{1}$ of the four-atom unit cell, the full $8\times8$ Hamiltonian $H_{C}(\mathbf{k})$ written in the $\Psi_{C}$ basis is periodic in $\mathbf{G}_{1}$ up to a unitary gauge transformation
\begin{equation}
H_{C}(\mathbf{k}+\mathbf{G}_{1})=D(\mathbf{G}_{1})\,H_{C}(\mathbf{k})\,D(\mathbf{G}_{1})^{\dagger},
\qquad
D(\mathbf{G}_{1})=\mathrm{diag}\bigl(e^{i\mathbf{G}_{1}\cdot\boldsymbol{\tau}_{i}}\bigr),
\label{eq:periodic}
\end{equation}
where $\boldsymbol{\tau}_{i}$ are the positions of the orbitals within the unit cell. Since $D(\mathbf{G}_{1})$ is unitary, $H_{C}(\mathbf{k}+\mathbf{G}_{1})$ and $H_{C}(\mathbf{k})$ are unitarily equivalent and have the same spectrum.

Under the change of basis $\Psi_{X}=U\Psi_{C}$ defined in Eq.~\eqref{eq:newbasis}, the Hamiltonian becomes block-diagonal [Eq.~\eqref{eq:hSBLGblock}],
\begin{equation}
H_{X}(\mathbf{k})=U H_{C}(\mathbf{k}) U^{\dagger}=
\begin{pmatrix}
h_{sg}(\mathbf{k}) & 0\\
0 & h_{sg}'(\mathbf{k})
\end{pmatrix},
\label{eq:HX}
\end{equation}
and the periodicity relation becomes
\begin{equation}
H_{X}(\mathbf{k}+\mathbf{G}_{1})=\widetilde{D}\,H_{X}(\mathbf{k})\,\widetilde{D}^{\dagger},
\qquad
\widetilde{D}=U D(\mathbf{G}_{1}) U^{\dagger}.
\label{eq:periodicX}
\end{equation}
The phase matrix $D(\mathbf{G}_{1})$ assigns the phases $e^{i\mathbf{G}_{1}\cdot\boldsymbol{\tau}_{i}}$ to the orbitals and distinguishes the $A,B$ orbitals from the $C,D$ orbitals, so that $\widetilde{D}$ is not block-diagonal in the two sectors: its off-diagonal blocks exchange $h_{sg}$ and $h_{sg}'$. This is the same statement as the single-layer relations $h_{s1}(\mathbf{k}+\mathbf{b}_{1}^{\mathrm{super}})=h_{s2}(\mathbf{k})$ and $h_{s1}'(\mathbf{k}+\mathbf{b}_{1}^{\mathrm{super}})=h_{s2}'(\mathbf{k})$: advancing $\mathbf{k}$ by one period turns the low-energy single-layer blocks into the high-energy ones. The same off-diagonal blocks of $\widetilde{D}$ act on the interlayer coupling, transforming $T_{\alpha,\beta}$ into $T_{\alpha,\beta}'$, which is the origin of the signs in Eq.~\eqref{eq:T1T2p}. Thus $H_{X}(\mathbf{k}+\mathbf{G}_{1})$ and $H_{X}(\mathbf{k})$ have the same spectrum, and the low-energy band structure of $h_{SBLG}(\mathbf{k})$, given by $h_{sg}(\mathbf{k})$, is the same as that of $h_{SBLG}(\mathbf{k}+\mathbf{b}_{1}^{\mathrm{super}})$, given by $h_{sg}'(\mathbf{k}+\mathbf{b}_{1}^{\mathrm{super}})$.

Bistritzer--MacDonald-like continuum models retain only the states near the Dirac point, so they are by construction restricted to the low-energy sector. In our model, however, merely enlarging the single-layer unit cell from the usual two atoms to four atoms already exposes part of the high-energy band structure. As shown in Eq.~\eqref{eq:HX}, this high-energy part can be decoupled and studied independently, exactly in the same way as the low-energy part; moreover, through the periodicity discussed above, it also carries the information of the low-energy sector. This observation suggests a natural direction for future work: to construct, along the lines of the present approach, a Bistritzer--MacDonald-like continuum model that applies to the high-energy sector as well, so that the entire band structure can be described in a unified way.

\end{document}